\documentclass[draft]{agujournal}
\drafttrue

\journalname{JGR-Planets}

\begin{document}

\title{Venus upper clouds and the UV-absorber from MESSENGER/MASCS observations}

\authors{S.~P\'{e}rez-Hoyos\affil{1}, A.~S\'{a}nchez-Lavega\affil{1}, A.~Garc\'ia-Mu\~noz\affil{2}, P.G.J.~Irwin\affil{3}, J.~Peralta\affil{4}, G.~Holsclaw\affil{5}, W.M.~McClintock\affil{5}, and J.F.~Sanz-Requena\affil{6}}

\affiliation{1}{Dpto. F\'{i}sica Aplicada I, ETSI, Universidad del Pa\'{i}s Vasco UPV/EHU, 48013 Bilbao, Spain}
\affiliation{2}{Zentrum f\"{u}r Astronomie und Astrophysik, Technische Universit\"{a}t Berlin, D-10623 Berlin, Germany}
\affiliation{3}{Department of Physics, Atmospheric, Oceanic and Planetary Physics, University of Oxford, Oxford, UK}
\affiliation{4}{Institute of Space and Astronautical Science (ISAS/JAXA), Sagamihara, Kanagawa, Japan}
\affiliation{5}{Laboratory for Atmospheric and Space Physics, Boulder, USA}
\affiliation{6}{Dpto. Ciencias Experimentales, Universidad Europea Miguel de Cervantes, 47012 Valladolid, Spain}

\correspondingauthor{Santiago P\'{e}rez-Hoyos}{santiago.perez@ehu.eus}




\begin{keypoints}
\item We analyze spectra of Venus equatorial atmosphere taken during the second MESSENGER flyby on 5 June 2007. 
\item Cloud tops are located at 75$\pm$2 km in the equatorial atmosphere.
\item Among all candidates proposed so far, S$_2$O and S$_2$O$_2$ provide the best agreement with the UV-absorption retrieved here.
\end{keypoints}

%
%


\begin{abstract}
One of the most intriguing, long-standing questions regarding Venus' atmosphere is the origin and distribution of the unknown UV-absorber, responsible for the absorption band detected at the near-UV and blue range of Venus' spectrum. In this work, we use data collected by MASCS spectrograph on board the MESSENGER mission during its second Venus flyby in June 2007 to address this issue. Spectra range from 0.3 $\mu$m to 1.5 $\mu$m including some gaseous H$_2$O and CO$_2$ bands, as well as part of the SO$_2$ absorption band and the core of the UV absorption. We used the NEMESIS radiative transfer code and retrieval suite to investigate the vertical distribution of particles in the Equatorial atmosphere and to retrieve the imaginary refractive indices of the UV-absorber, assumed to be well mixed with Venus' small mode-1 particles. The results show an homogeneous Equatorial atmosphere, with cloud tops (height for unity optical depth) at 75$\pm$2 km above surface. The UV absorption is found to be centered at 0.34$\pm$0.03~$\mu$m with a full width half maximum of 0.14$\pm$0.01 $\mu$m. Our values are compared with previous candidates for the UV aerosol absorber, among which disulfur oxide (S$_2$O) and dioxide disulfur (S$_2$O$_2$) provide the best agreement with our results.
\end{abstract}

%
%

%


%
%
%
%

\section{Introduction}

Venus is the only known terrestrial planet that is fully cloud-covered. These clouds are optically thick in the visible wavelengths, so seeing the surface requires observations at longer wavelengths to gaze through them. However, Venus clouds themselves have been studied for more than forty years \citep{venusI} and particularly since the arrival of the first space missions (Veneras and Pioneer Venus series). Since 1962 fifteen missions have visited the planet and provided a huge amount of information \citep{zasova2007}. However, many aspects are still missing, some of them quite important to fully understand Venusian atmospheric dynamics, chemistry and energy budget.

The vertical distribution of clouds is one of the aspects that we now understand better. Clouds are distributed roughly from altitudes of $\sim$ 50 km up to $\sim$ 70 km, although there are hazes that can be found both above and below such levels \citep{venusII}. We know since the 1970s that most of the visible light is scattered by spherical particles of 1 $\mu$m radius composed by a liquid mixture of water and sulphuric acid at 75 \% concentration \citep{venuspol}. This particle distribution is responsible for distinct features such as the primary rainbow and the glory that can be seen even in the disk-integrated reflecting properties of the planet \citep{venuspol,agm_glory}. Many recent works have reported the photometric or polarimetric glory observations on Venus clouds \citep{petrova,rossi,shalygina}. Such particles are accompanied by other statistical modes whose vertical distribution and overall latitudinal variation are well known in general terms, although systematic analysis of their temporal and spatial variation is still lacking.

The aspect of the Venus' clouds is bland and featureless in most wavelengths but, when observed in the near-ultraviolet, particularly at 0.36-0.37 $\mu$m, lots of structure emerges \citep{peralta} revealing the dynamics at Venus upper clouds. In fact, it was the UV observations of Venus \citep{ross} that revealed its westward atmospheric superrotation \citep{boyer}. A number of spacecraft and ground-based facilities have taken advantage of this spectral range to investigate Venus dynamics: Mariner 10 \citep{murray}, Pioneer Venus \citep{rossow}, Galileo \citep{belton} and Venus Express \citep{titov2012,khatuntsev,bertaux}, to name a few. In recent time, it has been possible to use modest aperture telescopes to provide support to JAXA's Akatsuki mission \citep{asl}. There is also some degree of cloud top contrast in near infrared images, mostly caused by different scattering properties below the cloud tops \citep{crispnir,takagi}.

The UV tracer is also discernible in the spectra of the planet, with a broad absorption band in the near-UV and blue side of the spectrum, roughly from 0.28 $\mu$m to 0.5 $\mu$m \citep{pollack}. Given its unknown origin, it is sometimes referred as the "mysterious UV-absorber". SO$_2$ was initially suggested as a candidate but it only contributes to short-wavelength absorption and its effect is negligible at wavelengths longer than 0.32 $\mu$m \citep{so2}, so another species is required. During the last decades, sulfur-bearing species have been often suggested as candidates for the unknown absorber \citep{pollack,toon} but an unambiguous identification is still missing. Very recently, disulfur dioxide has been proposed as the mysterious absorber \citep{frandsen}, in the form of the isomers $cis-$OSSO and $trans-$OSSO, as the calculation of their properties matches the spectral signature of the absorption and makes a plausible case for the chemistry required for their formation. Other authors have proposed different species over the years \citep{zasova1981} and some of these candidate species have been backed by detailed chemistry models \citep{krasnopolsky2016} that provide reasonable sinks and sources for the required products . As of today, there is no general agreement on the nature of the UV-absorber in Venus and thus this remains as one of the most intriguing open questions in planetary atmospheres \citep{krasnopolsky2006}. An excellent review of a good number of candidates proposed so far can be found in \cite{mills} and a short-list is given in Table \ref{table:candidates} in section~\ref{uvcandidates}.

Venus' clouds  and the UV-absorber also have a direct influence on the planetary energy budget. Even though Venus is closer to the Sun and receives more solar flux than the Earth, the thick cloud cover scatters and/or absorbs more than 50\% of it above 64 km \citep{tomasko}. This produces an intense heating of 8K/day \citep{crisp} that has been proposed to be one of the engines of the atmospheric superrotation through the excitation of thermal tides. Models based on Venus Express observations \citep{lee2015b} show that variations in the scale height of small aerosols create significant changes in the radiative forcing, with a vertical extension of the upper cloud layer implying a decrease of the outgoing thermal flux and enhancement of the mesospheric cooling. The global energy budget of the planet has been recently revisited by a complete reanalysis by \cite{haus2015, haus2016}. In any case, it is obvious that a good characterization of the absorption in the range from 0.32 $\mu$m to 0.5$\mu$m is essential if we want to understand the planetary energy budget and therefore Venus' dynamics.

The goal of this paper is to study the vertical distribution of particles in the mesosphere of Venus using data obtained during the second flyby of the MESSENGER spacecraft in route to Mercury. We will analyze near-UV to near-infrared spectra taken at the Equatorial region of the planet on 5 June 2007. We will particularly focus in the absorption properties of the mysterious UV-absorber in order to compare our results with currently proposed candidates for the absorption.

The paper is organized as follows. We describe the data in Section~\ref{data}, where we discuss the spatial and spectral behaviour of the measurements, as well as some cross-calibration issues between the visible and near infrared arms of the instrument. Section~\ref{methods} is devoted to a description of our methods, in particular the radiative transfer and retrieval technique, and a description of the a priori assumptions and the free and fixed parameters of the model. Results are presented in Section~\ref{results}, both regarding the general cloud properties and the UV-absorption. Such results are discussed in Section~\ref{discussion}, in terms of the vertical particle distribution, the UV-absorber candidates and an evaluation of the impact of the present results on the energy budget of Venus' atmosphere. Finally, we summarize our main conclusions in Section~\ref{conclusions}.

\section{Data} \label{data}
\subsection{MASCS/MESSENGER}

In this work we have used data collected by the Visible and InfraRed Spectrograph (VIRS) in the Mercury Atmospheric and Surface Composition Spectrometer (MASCS) instrument on board NASA's MErcury Surface Space ENvironment GEochemistry and Ranging (MESSENGER) mission. A complete description of the instrument can be found in \cite{mascs}. VIRS uses two different detectors for visible (VIS, 0.3 -- 1.05 $\mu$m) and near-infrared wavelengths (0.85 -- 1.45 $\mu$m), although there is a gap in the calibrated data that leaves no overlap between channels. Both channels have a resolution of 4.7 nm and a dispersion of 2.33 nm per pixel.

\begin{figure}[h]
\centering
\includegraphics[width=20pc]{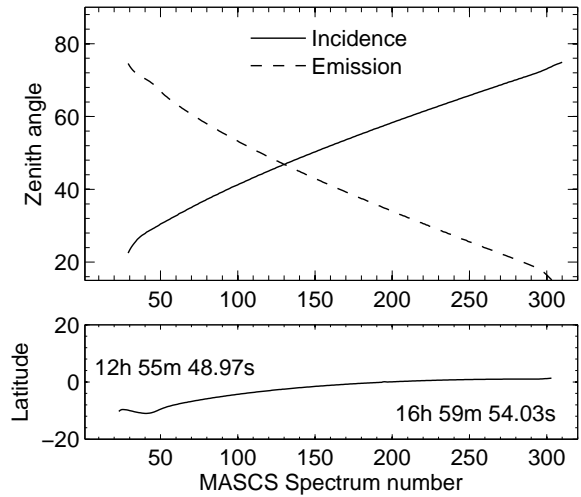}
\caption{ MESSENGER/MASCS observing conditions on June 5, 2007: (top) Incidence and emission zenith angles for all spectra analyzed here. (bottom) Mean latitude of each spectrum. Local times for the first and last spectrum are also indicated.}
\label{geom}
\end{figure}

MESSENGER performed its second Venus flyby in early June, 2007 \citep{messenger}. This was a gravity assist maneuver in order to put the spacecraft in an orbit closer to the Sun. During this process, the greatest acceleration of the mission was attained. VIRS data were acquired at a quite constant rate from 22:55 UTC to 23:02 UTC as MESSENGER moved from close to the sub-solar longitude (12:50:25 Local Time, longitude $-$88.72$^\circ$) to the terminator (16:54:34 LT, longitude $-$160.4$^\circ$) along Venus' Equator with a phase angle close to $\alpha \sim$ 90$^\circ$ $\pm$0.5$^\circ$. This implies a length  of the footprint of almost 7,500 km from the beginning to the end of the data acquisition. Figure~\ref{geom} shows the incidence and emission angles and the latitude footprint of the $\sim$ 300 usable spectra. Here we have removed a few spectra with extreme viewing or illumination angles (zenith angles $>$ 75$^\circ$) that will not accommodate the plane-parallel approximation of our radiative transfer model, described in section~\ref{methods}. Additionally, 12 spectra were removed due to low signal for unknown reasons, resulting in 318 spectra. The average size of the footprint on Venus cloud tops is $\sim$ 0.1$^\circ$ in latitude and $\sim$ 0.2$^\circ$ in longitude. Almost simultaneously, some images were also taken with the Mercury Dual Imaging System (MDIS) \citep{mdis} that will be used for VIS and NIR channels cross-calibration in subsection~\ref{cal}.

\subsection{Data overview}

The spectrum of Venus in our wavelength range is dominated by a few species. An illustrative review of the transmission spectrum of Venus, seen as a transiting exoplanet, can be found in \cite{venus_transiting}, particulary at figure 2, and \cite{agm2012}. Starting at the shortest wavelengths, SO$_2$ has a maximum absorption at around 0.3 $\mu$m and then decays rapidly towards 0.32 $\mu$m \citep{so2}, where its effect is mostly negligible. Immediately afterwards, there is a wide band ending well past 0.5 $\mu$m which is caused by the mysterious UV-absorber extending into the blue. The rest of the bands in our range are mostly caused by H$_2$O, in spite of its low abundance (see Table~\ref{table:parameters}), with the notable exception of the CO$_2$ band around $\sim$ 1.4 $\mu$m. This is a rather weak CO$_2$ band compared to others at longer wavelengths but the massive presence of this gas makes it a prevalent constituent in our spectra. CO$_2$ overlaps with water also at $\sim$ 1.4 $\mu$m (and, to a minor extent, at other wavelengths) but H$_2$O displays in general wider bands.

In the following, we will use for the measurements indistinctly the radiance (expressed for example in Wm$^{-2}$sr$^{-1}\mu$m$^{-1}$) and the reflectivity. Reflectivity $I/F$ \citep{aslbook} is defined as the ratio between the observed radiance and $\pi F_{\odot}$  where $F_{\odot}$ is the solar flux \citep{colina} at normal incidence at Venus' distance.

\begin{figure}[h]
\centering
\includegraphics[width=20pc]{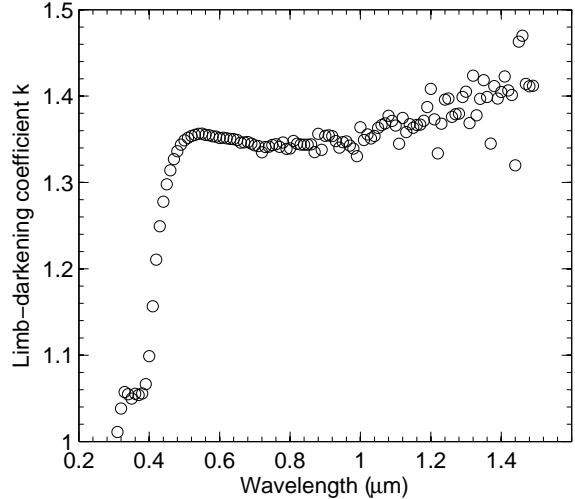}
\caption{Limb-darkening coefficient $k$ as a function wavelength.}
\label{minnaert}
\end{figure}

As a first approach to the data, we fitted the observed dependence of the reflectivity with the illumination and viewing angles by using a simple Minnaert law \citep{minnaert}. This empirical law reproduces the reflectivity as a function of the cosines of the incidence ($\mu_0$) and emission ($\mu$) angles:

\begin{equation} \label{eqminnaert}
	(I/F) = (I/F)_0 \mu_0^k \mu^{k-1}
\end{equation}
 
In equation~\ref{eqminnaert}, $(I/F)_0$ is the geometry corrected nadir-viewing reflectivity and $k$ is the limb-darkening coefficient. A Lambertian surface, for example, would have a value of $k = $ 1.

The data provide a good fit to this expression within the observed range of viewing and illumination conditions, and the resulting limb-darkening as a function of wavelength is shown in figure~\ref{minnaert}. It is interesting to note that there are three distinct regions. Starting from longer wavelengths, above 0.5 $\mu$m limb-darkening ranges $k = $ 1.35--1.40. There is a transitional part of the spectrum where the UV absorber dominates and $k$ is reduced down to values of 1.05. Finally, closer to the SO$_2$ absorption band, the limb-darkening coefficient reduces again and gets closer to Lambertian values and the observed radiance is diffuse, independent of the observation angle. This is valid for the Sun-Venus-spacecraft configuration at the moment of the flyby and would require more diverse observation conditions to be generalized. While MASCS cannot provide as much spatial resolution as an imaging device, this piece of information is very interesting when normalizing the radiance at a given wavelength to study the dependence on scattering angle, as done for example by \cite{petrova} and \cite{shalygina}. 

\begin{figure}[h]
\centering
\includegraphics[width=20pc]{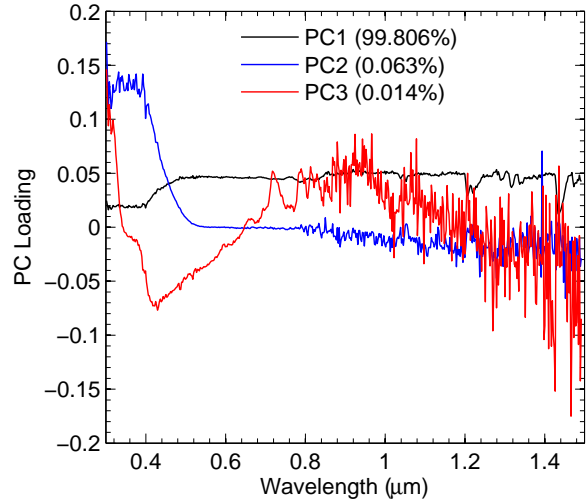}
\caption{Loadings of the three first principal components, which account for 99.883\% of the variance. The following PCs account for less than 0.002\% of the variance each.}
\label{pc}
\end{figure}

This analysis suggests that our data have a low spatial variability. In order to further investigate this aspect, we performed a Principal Component analysis (PCA, \cite{pca}). The PCA provides orthogonal (independent) contributions to the absolute variance and hence is able to disentangle the spectral contribution to spatial variability. Figure~\ref{pc} shows the spectral loading of each of the three first Principal Components. The first component PC1 accounts for more than 99.8 \% of the spatial variance of the MASCS spectra analyzed here. It resembles a common Venus spectrum (or the normalized average of the spectra) with the signature of SO$_2$ in the short wave side, the UV-absorber, and the H$_2$O and CO$_2$ bands included in this spectral range. This implies that all the variability affects all wavelengths equally and therefore it can be assumed that variability is due to geometrical effects alone and it is basically the limb-darkening presented previously. The second component PC2, has a broad signature that is well correlated with the expected UV-absorption. However, PC2 only accounts for 0.06 \% of the variance, and hence, even though this is the strongest isolated source of variability, its contribution to the total variance is very low. Component PC3 is even lower (0.014\% of the total variance) and has a very noisy spectrum in the near-infrared side, but it has a clear signature that could be related to the SO$_2$.

\subsection{Calibration issues}\label{cal}

The first challenge with VIRS data is that VIS and NIR sides of the spectrum have a cross-calibration bias. When radiances are transformed into reflectivity, the NIR side is clearly brighter than the VIS side. The calibrated spectra have a gap around $\sim$ 0.825 $\mu$m with no overlap between the channels \citep{asl}. If we compute the expected Venus spectrum by using the reasonable a priori model defined in section~\ref{apriori} we find that continuum at both sides of the gap should be mostly the same. There should be no discontinuity in the data as the aerosol and gas optical properties do not introduce such behavior. If we focus on the 0.570--0.670 $\mu$m region for the VIS continuum and the 0.970--1.030 $\mu$m in the NIR side, models show that the NIR continuum would be a 5\% darker at most, depending on the model parameters. At this point, it was possible to correct the VIS spectrum to match the NIR continuum or vice versa. Initial models favored the values observed at VIS wavelengths, and the NIR spectra were brighter than expected.

\begin{figure}[h]
\centering
\includegraphics[width=20pc]{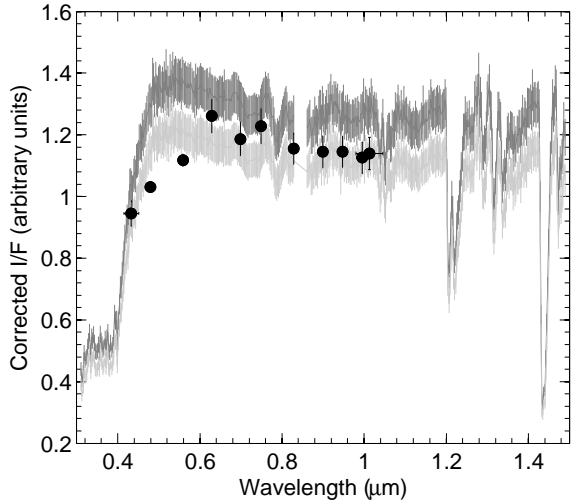}
\caption{Two alternative cross-calibrations of VIS and NIR sides of MASCS-VIRS spectra compared with average MDIS values (circles with error bars) taken almost simultaneously. In light grey, a correction of the NIR side to match the VIS continuum, the opposite is shown in darker grey.}
\label{mdis}
\end{figure}

\begin{figure}[h]
\centering
\includegraphics[width=20pc]{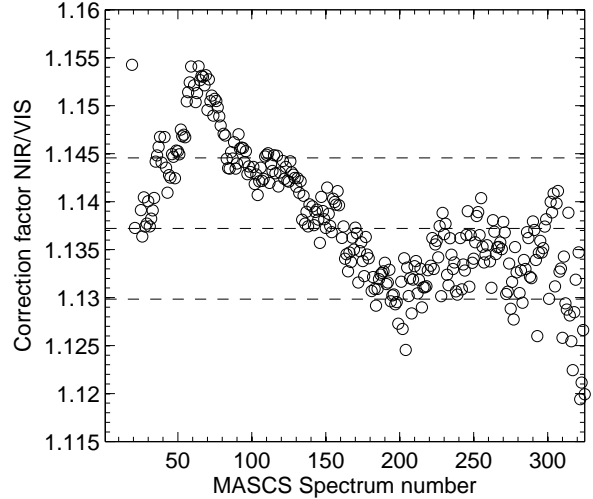}
\caption{Ratio of the NIR to VIS continuum for every MASCS spectrum. In average, NIR side is 13.7$\pm$0.7\% brighter than expected. Dashed horizontal lines show the average value and its 1-$\sigma$ standard deviation.}
\label{correction}
\end{figure}

In order to look for an independent confirmation, we navigated and calibrated almost simultaneous images obtained with the MESSENGER/MDIS instrument \citep{peraltaMDIS}. MDIS filters only covered the VIS side of MASCS data but were in agreement with that part of MASCS data. Figure~\ref{mdis} shows a comparison between both instruments averages, geometrically corrected following the method described above. While the agreement is not perfect, we decided to reduce the brightness of the NIR by a factor computed for each spectrum by evaluating the continuums described in the previous paragraph. The average correction factor was 1.137 $\pm$ 0.007 and its dependence with the spectrum number is shown in figure~\ref{correction}.

This correction also has an impact in the assumed error bars of the spectrum. Here, we have assumed a 5\% relative error in all spectra, with a minimum error in radiance of 1~mWcm$^{-2}\mu$m$^{-1}$srad$^{-1}$. This prevents an excessive weight of the lower radiance values (particularly at absorption bands) in the fitting process, relative to the continuum values. It should be noted that the error bars are not of excessive interest, as the retrieval technique is focused in the minimization of the deviation, not in its exact value. However, the relative weight of error bars can play a role in the retrieval results. 

Finally, there are some broad features in the spectra that have no exact counterpart in our modeling. These are located in 0.685--0.760 $\mu$m, 0.763--0.810 $\mu$m and 0.815--0.830 $\mu$m. Being at the end of the VIS range and, after testing some candidates to explain the absorption, no match was found and we assumed they were artifacts, similarly to the ones reported by \cite{shalygina} using VeX/VIRTIS. We found no agreement either between the features reported in that work and those seen here.

\section{Methods} \label{methods}
\subsection{Radiative transfer code}

We have used the radiative transfer and retrieval suite NEMESIS \citep{nemesis} in order to interpret the observations. NEMESIS stands for Non-linear optimal Estimator for MultivariatE spectral analySIS, and it is based on an optimal estimator scheme \citep{rodgers}. The radiative transfer calculations are made in the correlated-k mode from k-tables pre-computed from line data obtained at the HITRAN 2012 database \citep{hitran}. The radiative transfer solver is based on the doubling-adding scheme \citep{hansentravis} on a plane-parallel atmosphere. The version of NEMESIS used here accepts as free parameters those describing the vertical distribution of gases and particles as well as the imaginary refractive indexes of particles as a function of wavelength.

This is is the first time that NEMESIS has been used for the inversion of Venus' day-side scattered data at visible wavelengths. We tested initial results with an implementation \citep{agm} of discrete ordinates DISORT method \citep{disort} including multiple-scattering of Venusian particles and both codes agreed within a few percent. This NEMESIS implementation has been exhaustively tested with Venus night-side emission as shown in previous publications \citep{tsang2008, tsang2010, barstow}.

\subsection{Model atmosphere}\label{apriori}

Table~1 summarizes the parameters describing the atmosphere of Venus in our model. Even though the atmosphere is divided in 20 vertical layers for computational purposes, here we prefer to separate the contribution of the gas, the UV-absorber and each of the particle modes. In this table we show the fixed or initial value for each parameter as well as a reference supporting the choice.

The role of the gas if defined by its scattering and absorption properties. Rayleigh scattering cross-section as a function of wavelength was computed for a mixture of CO$_2$ and N$_2$ \citep{tsang2010}. In the wavelength range of VIRS data there are absorption bands from CO$_2$ and H$_2$O. Line data for them was retrieved from HITRAN 2012 database and transformed to k-tables. As already explained, the SO$_2$ plays a substantial role at the UV-side of the spectrum. Absorption cross-section values of SO$_2$ were taken from \cite{so2}. The abundance of the main consituents were taken from \cite{vira}. The variation of water abundance with altitude followed a simple model, with a high altitude value taken from \cite{fedorova} and a low altitude one from \cite{tsang2010}. Between 45 and 70 km values were interpolated linearly in altitude. The SO$_2$ abundance is known to vary with altitude \citep{fedorova,belyaev2012,belyaev2017} but initial runs showed this subtlety to have little impact on the model results, so we decided to use a constant SO$_2$ abundance.  We leave the analysis of the SO$_2$ vertical profile as a future work. In addition, since its abundance is highly variable with time, we used the values inferred by \cite{marcq2013} for 2007.

The particles require a more sophisticated description. All modes except mode-1 were assumed to be composed by an aqueous solution of H$_2$SO$_4$ concentrated at 75 \%, whose refractive index (real and imaginary parts) were taken from \cite{palmer}. We followed here the approximation of having the UV-absorber well-mixed with the smallest particles (mode-1), as done in other previous works \citep[e.g.][]{crisp}. This way, for mode-1 particles we only need the imaginary part of the refractive index to model the UV-absorption. The real part of the refractive index is computed using the Kramers-Kronig relation, taken as a reference the real refractive index of H$_2$SO$_4$ at 0.5 $\mu$m. As initial values, we took the results by \cite{pollack} for the same approximation.
 \begin{table} 
 \caption{Parameters of the atmosphere}
 \label{table:parameters}
 \centering
\begin{tabular}{c c c c c}
\hline 
{\bf Layer}    & {\bf Parameter}  & {\bf A priori} & {\bf Type} & {\bf Reference}  \\ 
\hline 
{\em Gas}  & vmr$(CO_2)$ & 0.965  & Fixed & \cite{vira} \\ 
           & vmr$(N_2)$  & 0.035  & Fixed & \cite{vira} \\ 
z $>$ 70km & vmr$(H_2O)$ & 1 ppm  & Fixed & \cite{fedorova} \\ 
z $<$ 45km & vmr$(H_2O)$ & 30 ppm & Fixed & \cite{tsang2010} \\ 
           & vmr$(SO_2)$ & 500 ppb & Fixed & \cite{marcq2013} \\ 
\hline
{\em UV-abs.}    & $m_i(\lambda)$ &  see Reference  & Free & \cite{pollack} \\ 
\hline
{\em Mode-1} & $z_1$(km) & 60    & Free & \cite{crisp} \\ 
           & $\tau_1$  & 4        & Free & \cite{crisp} \\ 
           & $H_1(H_g$) & 1    & Free           & \cite{tsang2010} \\ 
           & $r_1, \sigma_1$ & 0.3 $\mu$m, 0.44        & Fixed   & \cite{barstow} \\ 
\hline
{\em Mode-2}     & $z_2$(km) & 60    & Free & \cite{tsang2010} \\ 
           & $\tau_2$  & 8        & Free & \cite{tsang2010} \\ 
           & $H_2(H_g$) & 1    & Free           & \cite{tsang2010} \\ 
           & $r_2$, $\sigma_2$ & 1.0 $\mu$m, 0.25       & Fixed   & \cite{barstow} \\ 
           & $m_r$,$m_i(\lambda)$ & 75\% H$_2$SO$_4$      & Fixed   & \cite{palmer} \\          
\hline
{\em Mode-2'}     & $z_{2'}$(km) & 45    & Free & \cite{tsang2010} \\ 
           & $\tau_{2'}$  & 8        & Free & \cite{tsang2010} \\ 
           & $H_{2'}(H_g$) & 1    & Free           & \cite{tsang2010} \\ 
           & $r_{2'}$, $\sigma_{2'}$ & 1.4 $\mu$m, 0.21       & Fixed   & \cite{barstow} \\ 
           & $m_r$,$m_i(\lambda)$ & 75\% H$_2$SO$_4$      & Fixed   & \cite{palmer} \\          
\hline
{\em Mode-3}     & $z_3$(km) & 45    & Free & \cite{tsang2010} \\ 
           & $\tau_3$  & 9        & Free & \cite{crisp} \\ 
           & $H_3(H_g$) & 1    & Free           & \cite{tsang2010} \\ 
           & $r_3$, $\sigma_3$ & 3.65 $\mu$m, 0.25       & Fixed   & \cite{barstow} \\ 
           & $m_r$,$m_i(\lambda)$ & 75\% H$_2$SO$_4$      & Fixed   & \cite{palmer} \\          
\hline

\end{tabular} \end{table}

The particle modes are distinguished by the particle size and vertical distribution. Regarding size distributions, we opted to follow the values listed by \cite{barstow} for log-normal distributions which, in the end followed from the values by \cite{pollack1993}. In short, it is well known that at a given altitude range it is possible to find particles of various sizes \citep{venusII}. The micron-sized mode-2 particles scatter most of the radiation at visual wavelengths, but slightly larger particles (the so-called mode-2') can be found at deeper levels. The two other families are the sub-micron sized mode-1 and the larger particles of mode-3, which have radii of a few microns.

In order to describe the vertical distribution of such particles, we decided to follow the simplified description by \cite{tsang2010}. In this model, the distribution of particles is represented by three parameters: base altitude, scale height and peak abundance (reached at base). The scale height refers to its value at the cloud base and it will be given in terms of the gas scale height, which is $\sim$ 4 km at the cloud tops. This does not pretend to be a realistic description of the actual distribution of particles, but a useful parameterization that focus on the values that the model can constrain, as explained in section~\ref{strategy}. In Table~1 we have given the total optical thickness at 0.63 $\mu$m provided by each particle mode, instead of the peak particle density, which is the parameter in the model. Hereafter we will refer to the optical thickness at this specific wavelength, except stated otherwise, in order to simplify the comparison with previous work. Also, the particle extinction coefficient (particle number density multiplied by cross section) is the most sensitive parameter for the model.

We only depart from \cite{tsang2010} description of the vertical distribution of particles with mode-1. A number of forward evaluations of the a priori model demonstrated that the UV-absorber should be higher in the atmosphere in order to account for the radiance at short wavelengths, leaving the rest of parameters fixed as stated in Table~1. This agreed better with the model atmosphere by \cite{crisp}, so we put the a priori base altitude at z$_1 = $ 60 km.

In summary, Table~1 defines a forward model that can be evaluated. When doing so, we find that it is not far from the actual data at certain geometries, particularly at intermediate illumination and viewing angles, but still requires some fine tuning that will provide useful information on the atmospheric parameters. But first, we must define which parameters are to be left as free and which ones fixed.

\subsection{Free parameters and fitting strategy} \label{strategy}

At this point, we need to evaluate which of the parameters that describe the atmosphere can be fitted with the current data. For doing so, we performed a number of forward evaluations, sample retrievals and computed derivatives in order to understand the a priori sensitivity. In order to have a complete description of the model sensitivity as a function of height and wavelength, we sometimes used a continuous vertical profile with 1 km vertical resolution, instead of the parameterizations described in the preceding section.

\begin{figure}[h]
\centering
\includegraphics[width=20pc]{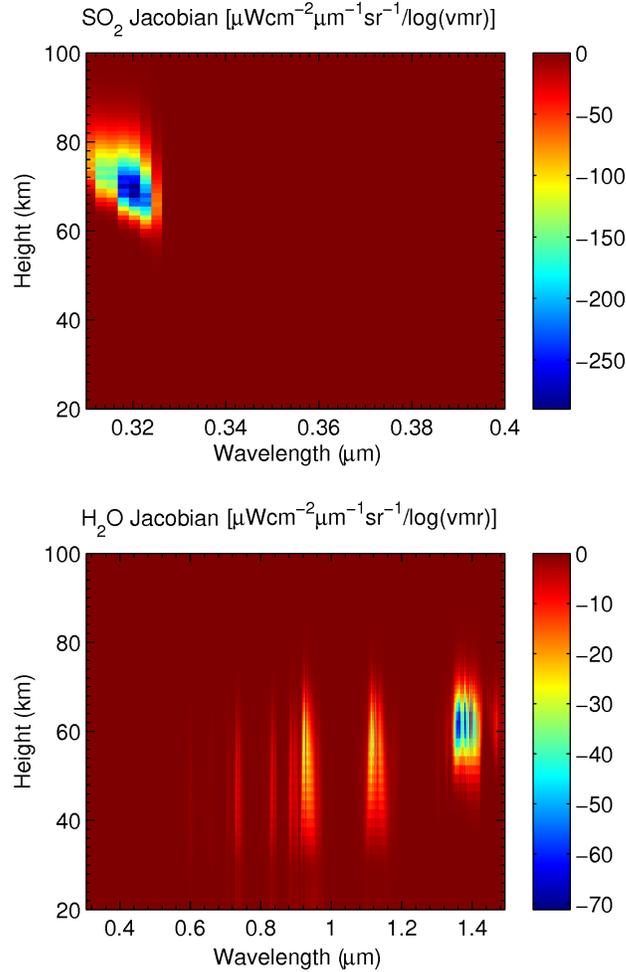}
\caption{Jacobians for (top) SO$_2$ and (bottom) H$_2$O as a function of altitude and wavelength. Please note the restricted wavelength range used for the display of SO$_2$ sensitivity. Units are given in terms of radiance change per volume mixing ratio change in logarithmic units.}
\label{gas_jacobian}
\end{figure}

Our initial idea was to include SO$_2$ and H$_2$O as free parameters. We show in Figure~\ref{gas_jacobian} the Jacobian matrix for both species (i.e., the matrix of partial derivatives of the radiance with respect to the parameter value at each altitude level). For this calculation, we assumed a constant abundance throughout the atmosphere. In the case of SO$_2$ we are only sensitive to concentration at altitudes 75$\pm$5 km at 0.32 $\mu$m. Recent works \citep{vandaele} have shown that the vertical distribution of SO$_2$ above the cloud tops can be very complex, even with an inversion layer at around 70--75 km. However, our sensitivity is narrow enough to support that we are only being sensitive to SO$_2$ abundance at the cloud tops, even though this parameter is going to be strongly coupled with the abundance of mode-1 particles and the UV-absorption in the 0.30--0.32 $\mu$m range.A future work on the SO$_2$ vertical profile could help to break this parameter degeneracy.

In the case of water, we have a range of sensitivities at different bands, mostly concentrated at 60$\pm$10 km (particularly at 1.4 $\mu$m), but with contributions for levels as deep as 40 km and even less where the absorption band becomes weaker. This supports the need to include a better description of water with altitude, as shown in Table~1. However, sample retrievals showed very low sensitivity to both species, with dispersions and error bars larger than 200\%, so we decided to fix the value of these parameters to reasonable values found in the literature, as already discussed.

In order to test the sensitivity to particle modes, we used a particle number density constant with height, instead of the parameterization presented above. This is not a realistic description of the vertical distribution, but it avoids abrupt cuts at cloud base.

\begin{figure}[h]
\centering
\includegraphics[width=18pc]{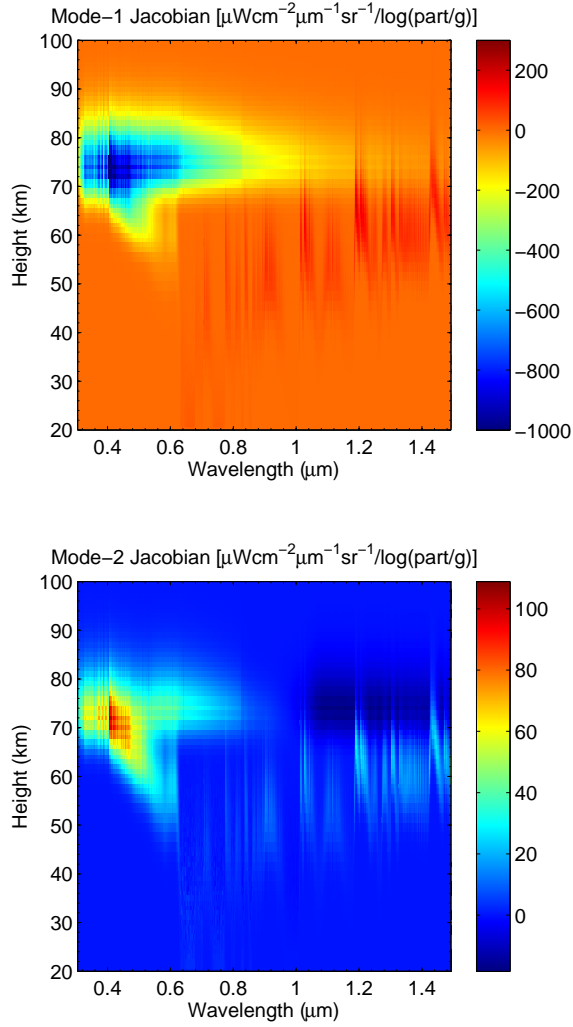}
\caption{Jacobians for mode-1 and mode-2 particles as a function of altitude and wavelength.}
\label{modes_jacobian}
\end{figure}

The model is most sensitive to the distribution of mode-1 and mode-2 particles. Mode-1 is dominant at short wavelengths, where the cross section is the largest due to particle size and to having the UV-absorption attached. In the near infrared bands, mode-2 has also a strong influence. Very roughly, since the figures depend on the exact description of each cloud layer, our models are sensitive to particles in the 50--80 km range, at most, and mostly at 66$\pm$6 km in the NIR bands and slightly higher, 75$\pm$7 km for the UV-absorption. It must be noted that our parameterization of particle distribution is defined for all atmospheric levels, while the retrieval is not necessarily reliable for all of them. Most of the radiance reflected by the planet in this wavelength range comes from the vertical levels described above, so the information on particle density above 80 km and below 50 km cannot be trusted and only the particle densities in the range of confidence can be taken as a robust conclusion of this analysis.

Regarding the simplified vertical description given in Table~1, we will see later that sensitivity is good for mode-1 and mode-2 parameters, except for their scale height. Mode-2' and mode-3 parameters are more difficult to constrain. In order to evaluate the sensitivity of each parameter from the actual retrievals, we used an improvement factor, as defined by \cite{improvement}:

\begin{equation} \label{eqimprovement}
	F(\%) = 100 \times (1 - \frac{\delta_{fit}}{\delta_{apr}} )
\end{equation}

In this equation, $\delta$ is the relative error of the a priori assumption and the fit as provided by the optimal estimator. We assumed a relative error of 25\% for all parameters except for the imaginary refractive index, which is 50\%. This way, an improvement factor $F = $ 0\% implies that our a posterior relative error is the same as the prior, and therefore contains less information than a retrieval with an improvement factor of $F = $ 100\%, whose relative error has been reduced drastically. The actual values for improvement factors, which will justify the election of free parameters, will be given in section~\ref{results}.

In order to avoid over-fitting the imaginary refractive index, we decided to split the wavelength range into two overlapping sub-ranges. This way, we first fitted wavelength above 0.5 $\mu$m fixing the mode-1 imaginary refractive index and leaving free the parameters of the vertical distribution of particles (12 free parameters in total). Then, the result is used as an input to fit wavelengths below 0.6 $\mu$m but now leaving free the imaginary refractive index (20 free parameters). This provides the best solution closest to our a priori for the UV-absorption.

\begin{figure}[h]
\centering
\includegraphics[width=20pc]{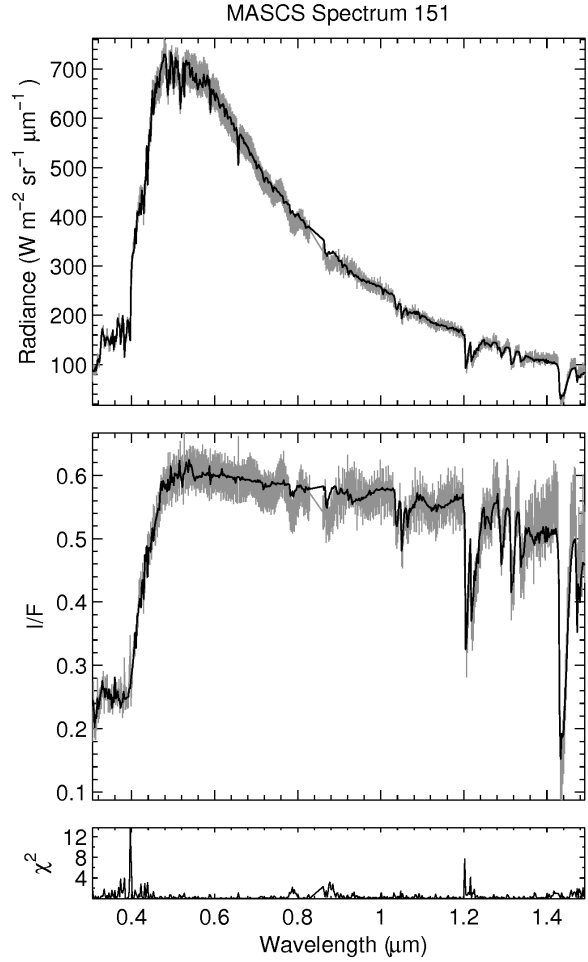}
\caption{Sample fit for MASCS spectrum 151 expressed in (top) radiance, (middle) reflectivity. Error ranges are shown in grey and best fit model is shown in black. The $\chi^2$ as a function of wavelength is also shown (bottom).}
\label{fit}
\end{figure}

\section{Results} \label{results}

\subsection{General comments}

The strategy summarized above provides a good fit to the data, with a few exceptions. Figure~\ref{fit} shows an example of fitting for an intermediate value of the illumination and viewing angles. The $\chi^2$ value is used as a diagnostic of the goodness of the fit, it is computed as the mean quadratic deviation in terms of the error bar of the data. With respect to wavelength, the most difficult regions is the slope around 0.4 $\mu$m, where higher resolution in the imaginary refractive index would provide a better fit. The next critical region is the absorption band at 1.2 $\mu$m, where fits are in general poorer. Finally, there seems to be a systematic offset between models in the red end of the spectra, as near infrared  continuum tends to be spectrally flatter than the models predict. We have tried non-systematic search of alternative models with a lower slope in the near infrared with negative results. This might be a calibration issue with the spectra, maybe because of differences in the solar spectrum used \citep{colina}.

\begin{figure}[h]
\centering
\includegraphics[width=20pc]{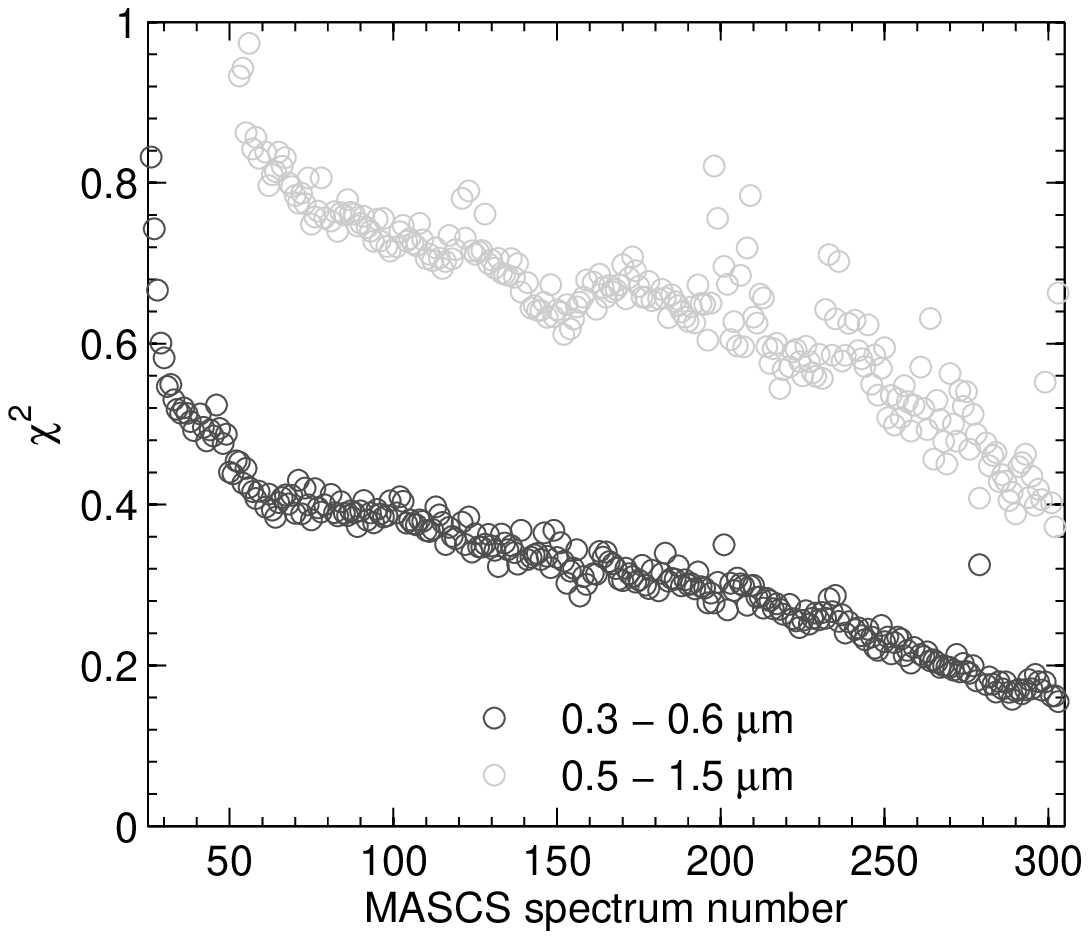}
\caption{Mean quadratic deviation for the VIS and NIR sides of the spectrum as a function of spectrum number.}
\label{chi2}
\end{figure}

In terms of the spectrum number, the first 50 spectra are problematic. They do not converge to low values of the mean quadratic deviation (shown in Figure~\ref{chi2}). This is particularly important in the near infrared fitting, since fitting the ultraviolet and visible is easier when the imaginary refractive index is left as a free parameter. Such spectra are coincidentally those with higher emission angles, as shown in Figure~\ref{geom}. While we only filtered zenith angles greater than 75$^\circ$, this suggests that the plane parallel approximation breaks at some 70$^\circ$. The model and its approximations (mainly the plane-parallel approximation) behave better at low emission angles. All in all, results for spectrum number below $\sim$ 60 are most likely not reliable.

\subsection{Cloud properties}\label{clouds}

\begin{figure}[h]
\centering
\includegraphics[width=30pc]{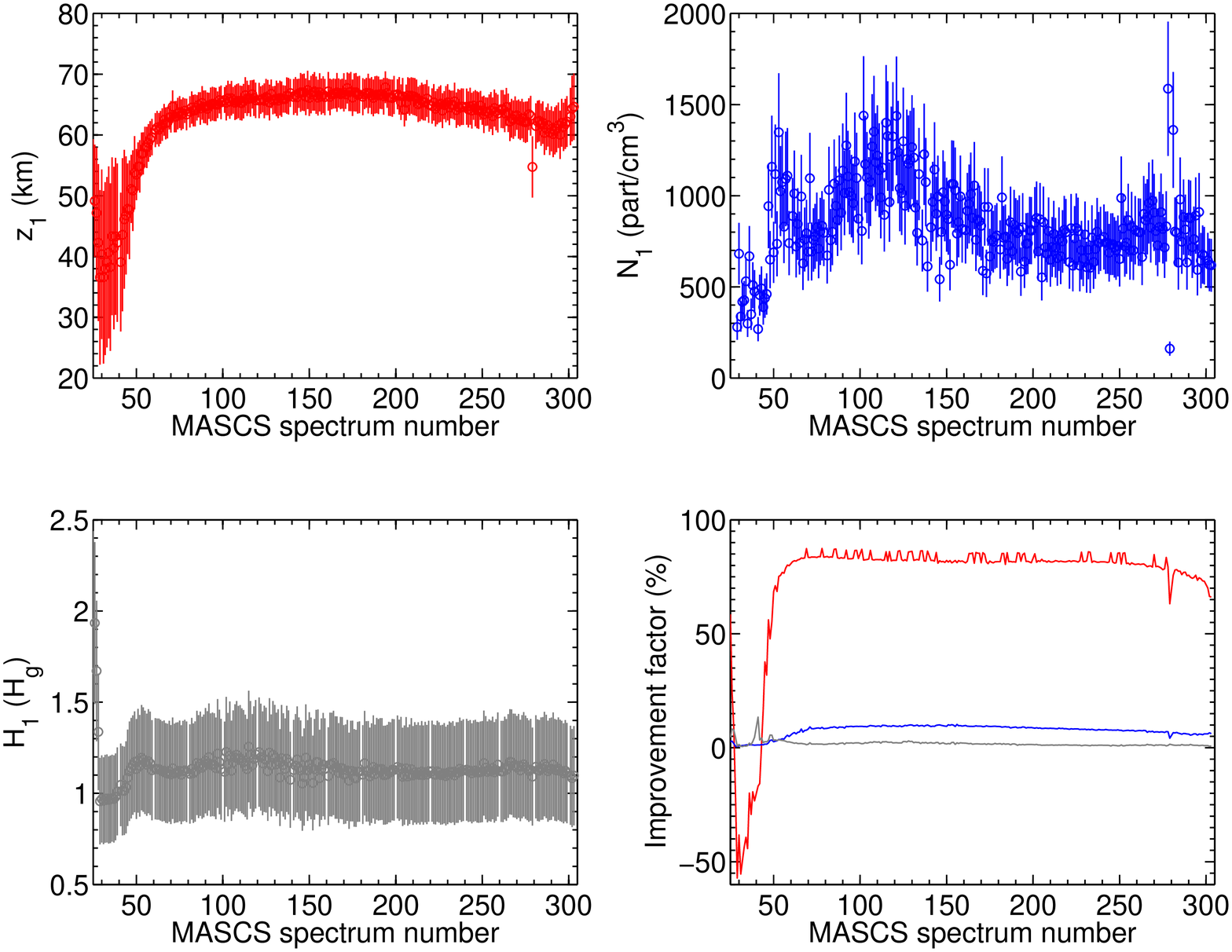}
\caption{Model results for mode-1 particles parameters and improvement factor. Red line is used for the cloud base height $z_1$, blue line for the particle number density $N_1$, and grey for fractional scale height $H_1$, the scale height in terms of the gas scale height $H_g$.}
\label{mode1}
\end{figure}

\begin{figure}[h]
\centering
\includegraphics[width=30pc]{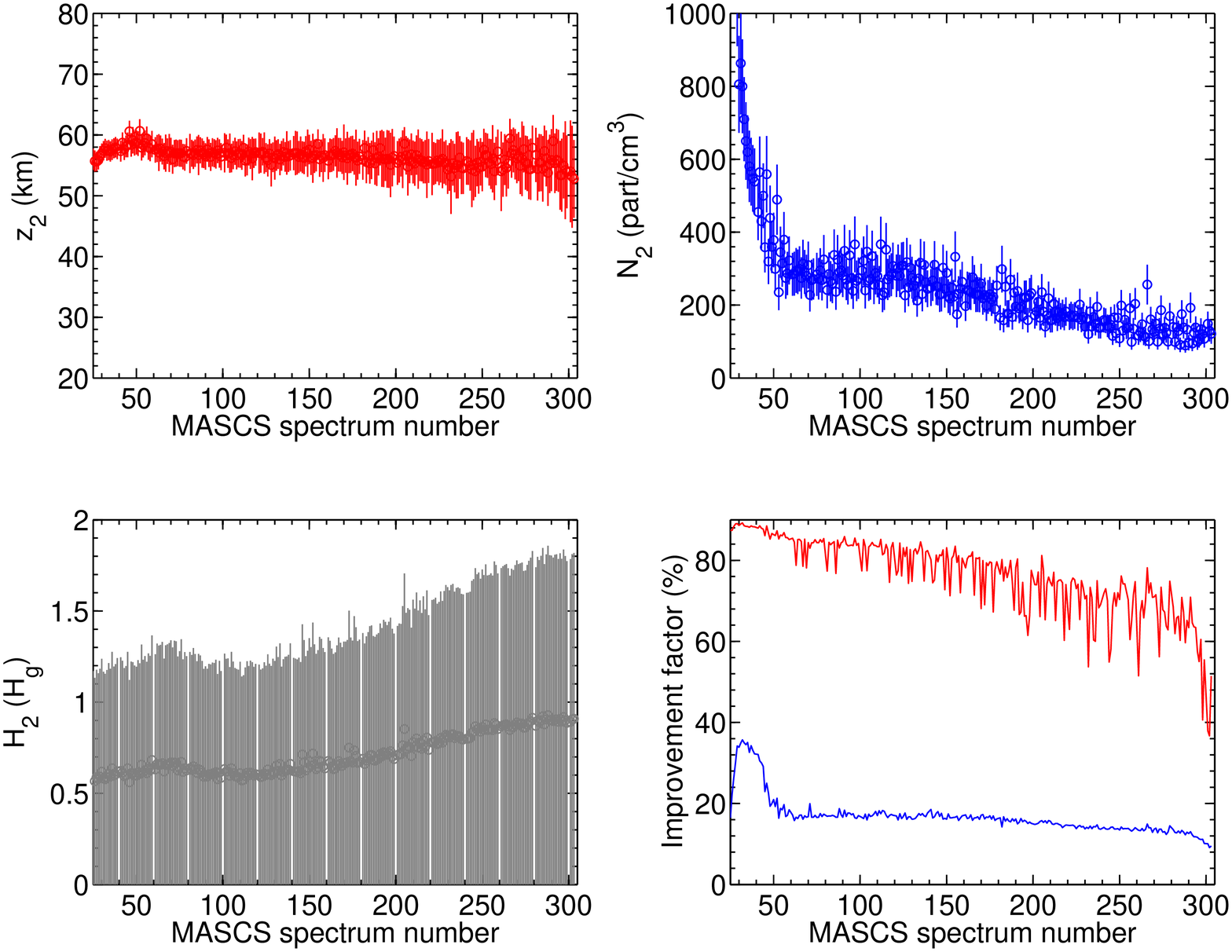}
\caption{Model results for mode-2 particles parameters and improvement factor.  Red line is used for the cloud base height $z_2$, blue line for the particle number density $N_2$, and grey for fractional scale height $H_2$, the scale height in terms of the gas scale height $H_g$.}
\label{mode2}
\end{figure}

\begin{figure}[h]
\centering
\includegraphics[width=30pc]{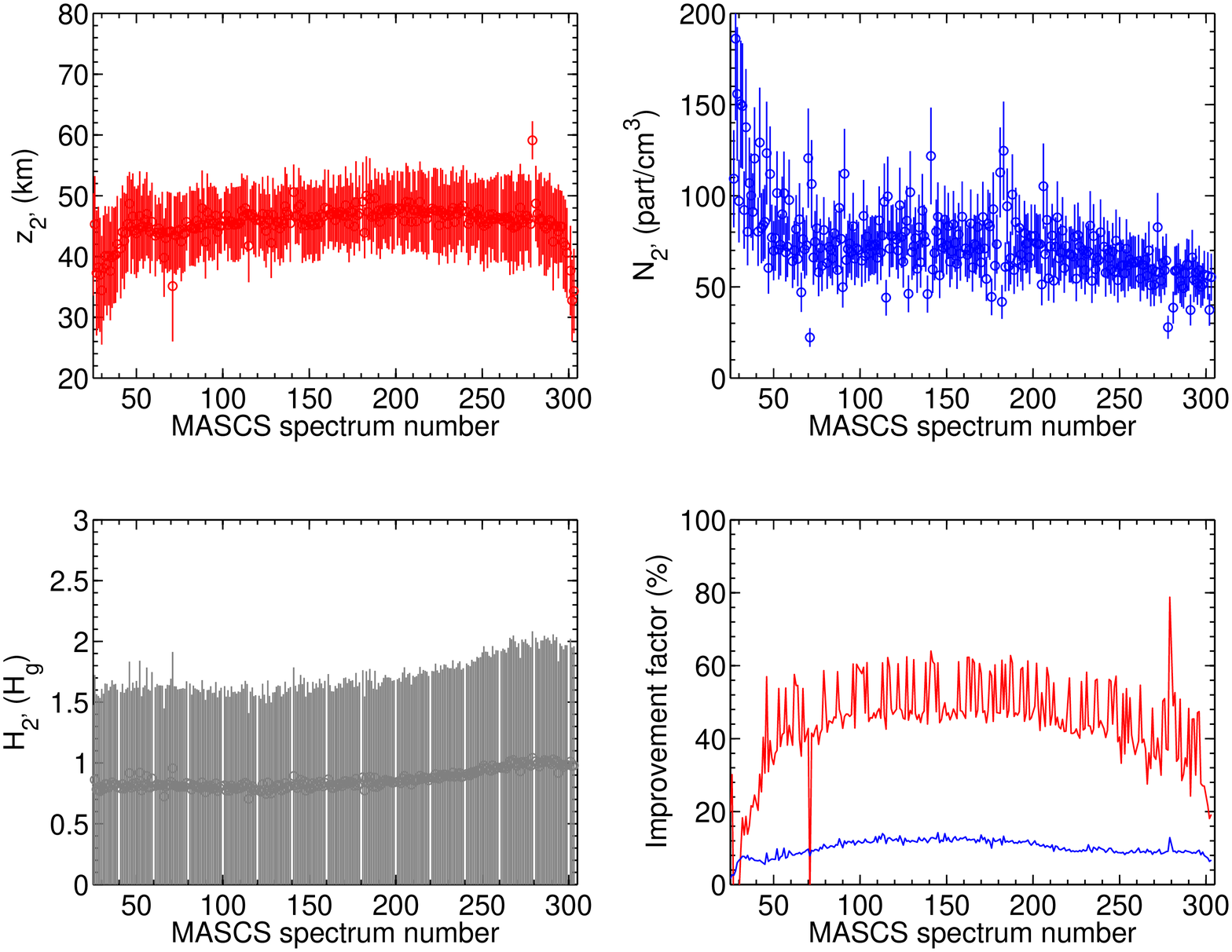}
\caption{Model results for mode-2' particles parameters and improvement factor.  Red line is used for the cloud base height $z_{2'}$, blue line for the particle number density $N_{2'}$, and grey for fractional scale height $H_{2'}$, the scale height in terms of the gas scale height $H_g$.}
\label{mode2p}
\end{figure}

\begin{figure}[h]
\centering
\includegraphics[width=30pc]{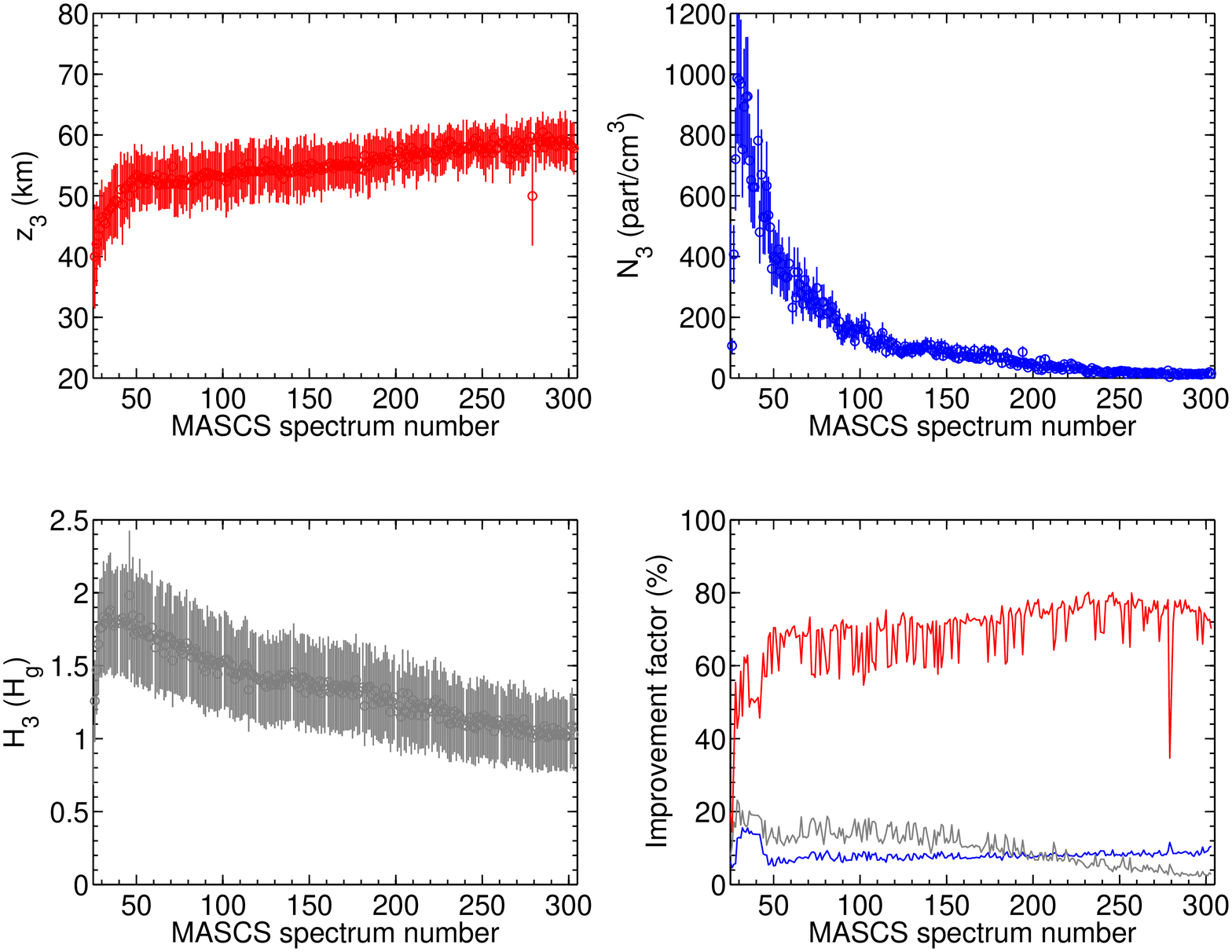}
\caption{Model results for mode-3 particles parameters and improvement factor.  Red line is used for the cloud base height $z_3$, blue line for the particle number density $N_3$, and grey for fractional scale height $H_3$, the scale height in terms of the gas scale height $H_g$.}
\label{mode3}
\end{figure}

Figures~\ref{mode1} to \ref{mode3} show the results for the vertical distribution of each of the particle modes. All figures show the parameters (base height $z$, particle peak density $N$ in part cm$^{-3}$, and scale height $H$ in terms of the gas scale height) and the improvement factor expressed as in equation~\ref{eqimprovement}.

Excluding spectra below number 60, Mode-1 particles (see Figure~\ref{mode1}) have a mean base height of 65$\pm$2 km and maximum particle number density of 900$\pm$200 part cm$^{-3}$, with a scale height of 1.14$\pm$0.03 $H_g$ ($\sim$ 4 km). The improvement factor is above 80\% for the altitude, implying that the relative error has been reduced from the 25\% assumed a priori to a mere 5\%. The particle density and the scale height show no improvement, while for the latter the dispersion of results is lower and values are closer to the a priori. The integration of the vertical profile, provides a mean optical thickness of $\tau_1$ = 3.2 $\pm$ 0.2 at 0.63 $\mu$m, notably lower than the initial value listed in Table~1.

Results for mode-2 (figure~\ref{mode2}) show the base of the cloud at 56$\pm$1 km with a more variable particle density of 200$\pm$70 part cm$^{-3}$. The scale height $H_2$ retrieval has a huge uncertainty but all models converge to values substantially lower than the prior (from 1 $H_g$ to 0.7 $H_g$ for the best fits, with a dispersion of only 0.1 $H_g$). The error in z$_2$, as in mode-1, has been strongly reduced and there is also some improvement in the particle density (average improvement factor of 15\%). The variable number of particles results in change of the total optical thickness from $\tau_2 \sim$ 10 for MASCS spectra around 50 down $\tau_2 \sim$ 5.5 for the last ones, which is a difference of almost a factor of 2.

The distribution of particles at deeper levels is more uncertain, as expected from the sensitivity analysis discussed above. Note that vertical levels above 80 km and below 50 km did not give enough sensitivity to provide a robust retrieval of the particle number density. Mode-2' cloud base is located on average at 46$\pm$2 km with a peak density of $N_2$ = 70$\pm$20 part cm$^{-3}$. However, it is not expected that cloud base can be located at such deeper levels with high temperatures \citep{holes}. Results for the scale height are inconclusive. Most likely, the model is unable to separate the contribution of mode-2' particles from that of mode-2 particles, as the particle sizes are very similar. This wavelength range does not look therefore adequate to discriminate between both modes. The integration of the particle density provides a mean optical thickness for mode-2' of $\tau_{2'}$ = 6.8 $\pm$ 0.4, with no substantial dependence on spectrum number.

However, the contribution from mode-3 particles, which are significantly higher than mode-2 or mode-2', seems to be more important. This is the only mode that displays a continuous variation along MESSENGER footprint, from the local noon to the evening terminator. The base height increases from 40 km to 60 km, while the average value is $z_3$ = 55$\pm$2 km. At the same time, the particle number density is reduced drastically from a maximum value of $\sim$ 1000 part cm$^{-3}$ to a few 10s of part cm$^{-3}$ (average value of N$_3$ = 70$\pm$60 part cm$^{-3}$). The scale height also changes from $H_3$ $\sim$ 2 H$_g$ to 1 H$_g$ (5--10 km) and an improvement factor around 20\%. When combining these values in an integrated measure, the total optical thickness of mode-3 particles is fairly stable with a mean value of $\tau_3$ = 7.5$\pm$0.4. This suggests that more mode-3 particles are required in the higher atmosphere where our retrieval is the most sensitive, while it differs in the way of making this possible. At 70 km, for example, the particle concentration of mode-3 particles is $N_3$(70km) = 1.6$\pm$0.3 part cm$^{-3}$, with slightly higher values at the last spectra, but low dispersion in general.

\subsection{UV-absorption}

\begin{figure}[h]
\centering
\includegraphics[width=20pc]{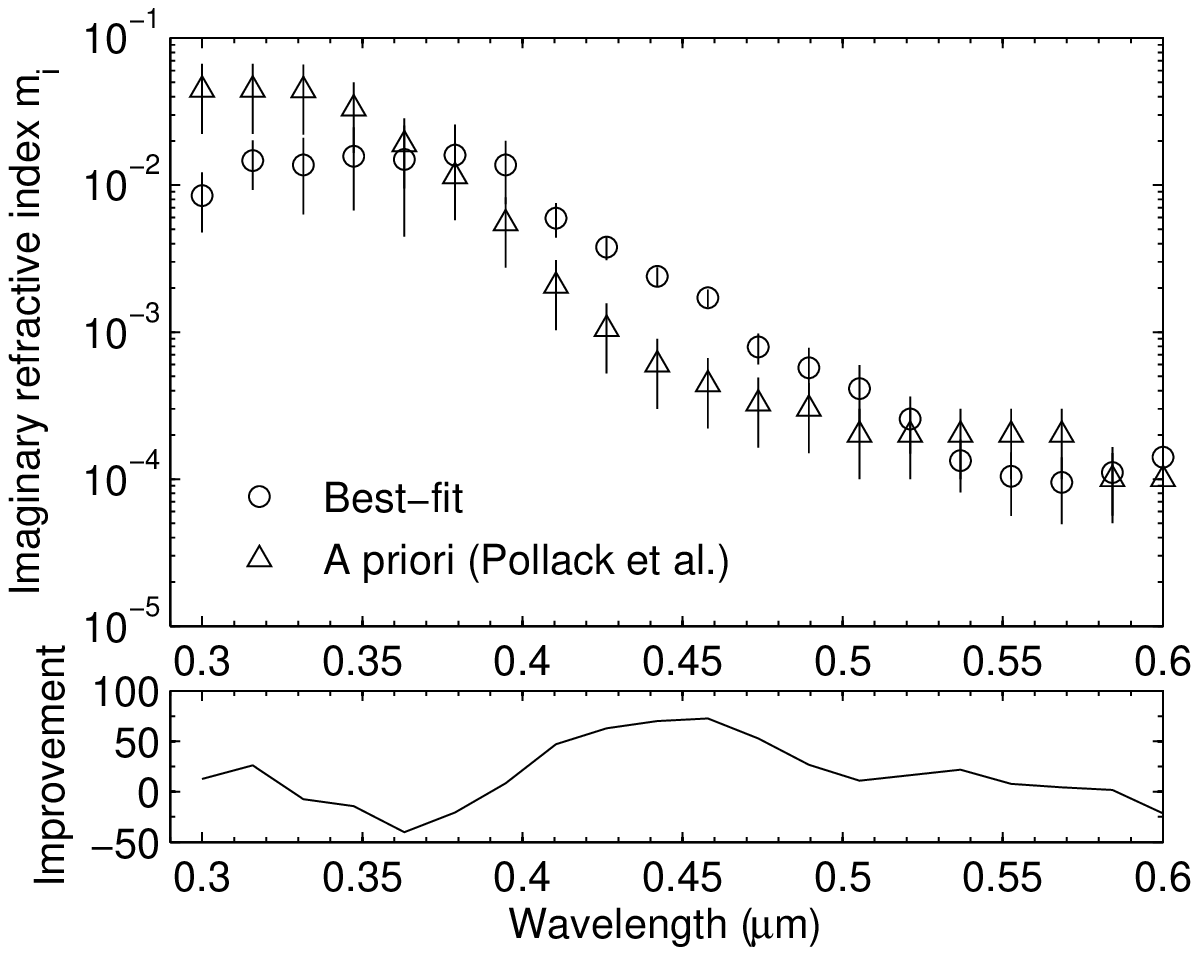}
\caption{Results for average imaginary refractive index $m_i$ compared with a priori assumption from \cite{pollack} and improvement factor.}
\label{mi}
\end{figure}

We show in Fig.~\ref{mi} the average results for the imaginary refractive index of mode-1 particles that fits the UV-absorption. The real refractive index is computed from the reference values and fitted imaginary values using the Kramers-Kronig relations, as discussed in section~\ref{apriori}. The resulting values of real part do not depart significantly from the initial values, less than 1\%. Other works \citep{petrova,rossi,shalygina} were better suited for determining the real refractive index and found similar values.

The initial values of the imaginary refractive index by \cite{pollack} can be described in terms of central wavelength absorption at 0.33 $\mu$m with a full width half maximum slightly below 100 nm, if we assume that the absorption band follows a perfect gaussian shape. These numbers must be taken cautiously as we only have information at one side of the absorption band and the few wavelengths in the short side might be affected by the SO$_2$ gas absorption as well.

Our results show a similar absorption to the prior derived by \cite{pollack}, but clearly displaced towards longer wavelengths, with maximum absorption at 0.34 $\mu$m and wider wings with FWHM = 140 nm, again assuming that the absorption band is gaussian. In figure~\ref{mi}, we show  the standard deviation of the values at each wavelength for all spectra, but retrieval errors are very similar in magnitude. As seen, results are more precise in the region from 0.4 to 0.45 $\mu$m, where the albedo rapidly increases for the central part of the absorption band. Values at the shortest wavelengths, where the absorption is the strongest, are more affected by the description of mode-1 particles (particle number density and height). With respect to the behavior at longer wavelengths (greater than 0.6$\mu$m, the results suggest that the tail of the absorption continues. However, the difference in absorption is so low with such values of the imaginary refractive index that we cannot robustly conclude that this is the case. The tail of the absorption in the red and near-infrared side of the spectrum deserves further research as it can help to elucidate the origin of the UV-absorption.

The retrieval differs significantly from the a priori values both at wavelengths below 0.35 $\mu$m (where our absorption is weaker) and in the region from 0.4 to 0.5 $\mu$m, where our particles absorb more strongly. This change in the particle absorption will be of interest when we discuss the nature of the UV-absorber in the following section.

Our results are in good agreement with independent determinations such as \cite{lee2017}, where the most probable values of the imaginary refractive index were evaluated by fitting observations of the glory with Akatsuki/UVI 283-nm filter. In the simulations, the peak of the glory became clearer with values similar to the largest ones found in this work, and the bottom of the UV-absorber located at similar altitudes. 

\newpage
\section{Discussion} \label{discussion}
\subsection{Vertical particle distribution}

A crude summary of the results presented in the preceding section can be found in Table~\ref{table:retrievals}. As long as our results do not depart much from the priors with respect to the vertical distribution of cloud particles, there is an obvious general agreement with the references they were based on. However, there are also a number of differences that are worth commenting on. First, our initial runs required the mode-1 to be located substantially higher than in the work by \cite{barstow}, with the mode-1 cloud base above 60 km rather than below 50 km. This is more in consonance with the description by \cite{crisp}, as it has been already discussed. Second, the cloud base for our mode-3 particles tends to be located higher in the atmosphere than in any previous work. This possibly implies the need of having more large-sized particles at higher altitudes (70 km), where our model is more sensitive. This could have been alleviated by incorporating absorption also on mode-2 particles, something that we discarded as an initial assumption but that could deserve further research.

\begin{table} 
 \caption{Summary of the retrievals}
 \label{table:retrievals}
 \centering
\begin{tabular}{c c c c c}
\hline 
{\bf Layer}    & {\bf Parameter}  & {\bf A priori} & {\bf Best-fitting}\\ 
\hline 
{\em Mode-1} & $z_1$(km) & 60    & 65$\pm$2\\ 
           & $\tau_1$  & 4        & 3.2$\pm$0.2 \\ 
\hline
{\em Mode-2}     & $z_2$(km) & 60    & 56$\pm$1 \\ 
           & $\tau_2$  & 8        & 10-5 \\ 
\hline
{\em Mode-2'}     & $z_{2'}$(km) & 45    & 46$\pm$2 \\ 
           & $\tau_{2'}$  & 8        & 6.8$\pm$0.4 \\ 
\hline
{\em Mode-3}     & $z_3$(km) & 45    & 55$\pm$2 \\ 
           & $\tau_3$  & 9        & 7.5$\pm$0.4 \\ 
\hline

\end{tabular} \end{table}

We have computed the cloud top as the level at which the one-way total optical depth is $\tau$ = 1 for wavelength of 0.63$\mu$m, as in previous works. In spite of the dispersion of particular values for each particle mode, the value is remarkably constant among all spectra with a mean altitude of $z_{top}$ = 75$\pm$2 km. This is very similar to the value of $\sim$ 74 km given by \cite{ignatiev} for the equatorial latitudes. The result by \cite{lee2012} is lower ($\sim$ 67$\pm$2 km) but it must be noted that this value is given at a longer wavelength (4.5$\mu$m) and that only mode-2 particles were used in that model. The combination of both facts necessarily results in a cloud-top at deeper level. However, \cite{lee2012} finds a particle scale height similar to the gas scale height as we did.

One of the best references with respect to the vertical distribution of particles 
in the Venusian atmosphere is the work done with the Pioneer Venus particle size 
spectrometer experiment, described by \cite{knollenberg}. 
There is an overall agreement with the direct measurements of the Pioneer Venus probe, 
with smaller mode-1 particles showing abundances in the 100s and larger mode-3 particles 
only in the 10s of particles per cubic centimeter. There also a notable agreement with 
the mode-1 total optical thickness  ($\tau_1$ = 3.2$\pm$0.2 in our work, 
versus 3.23 in \cite{knollenberg}, table 5. However, the mode-2 in \cite{knollenberg} 
accounts for an optical thickness of 9.76, while the combination of our mode-2 and mode-2' 
is above $\tau_{2+2'} >\sim$ 12 for most spectra. This is compensated 
by mode-3 particles, which in our results only account for half of the total 
optical thickness as they did in \cite{knollenberg}. However, we are not very sensitive 
to the distribution of mode-3 particles and the agreement with direct 
measurements is good in the levels around 65 km.

Regarding more recent results, comparing with the results by \cite{molaverdikhani} we find that the number of mode-1 particles is about the same above the cloud base (located at 60 km in their work), although they retain some particle abundance down to 48 km, something that we do not include in our model. There are some discrepancies though, as they assume smaller particles for mode-1 and thus the mode-1 opacity should be lower in their model. In the case of mode-2 particles we have similar figures of the order of a few 100s of part cm$^{-3}$ but they are possibly higher in this work for most cases. This happens again with mode-3 particles: similar order of magnitude but slightly higher for our results. The integrated aerosol opacity in our case is around $\tau \sim$ 25 at 0.63 $\mu$m, which seems close enough to typical estimations of the total particle load in the atmosphere \citep{venusII}.

In the case of the hazes located in the upper atmosphere, works by \cite{gao} and \cite{luginin} dealt with the distribution of particles above the main cloud deck. Our results compare well with those by \cite{luginin} at 75 km, where we find a mean of $\sim$ 100 part cm$^{-3}$ of mode-1 particles and $\sim$ 1.5 part cm$^{-3}$ of mode-2. This is comparable with their 500 part cm$^{-3}$ and 1 part cm$^{-3}$, respectively. At 90 km, however, the agreement is poorer and we get about one order of magnitude fewer particles for both modes. As our sensitivity at those levels is very low, this difference is not conclusive and their values should be trusted.

All in all, the average vertical distribution of particles is typical of the equatorial region of Venus, with cloud-tops higher than what is commonly retrieved at higher latitudes \citep{ignatiev}. For the levels between 50 and 70 km, where we are the most sensitive, the results yield no surprises.

\subsection{UV-absorber candidates}\label{uvcandidates}

There are a number of issues that must be studied in order to constrain the nature of the UV and blue absorber longwards of 0.32$\mu$m. First, any proposed candidate (or combination of them) should match the spectral signature of the UV-absorber. Second, the expected number density of the candidate should be consistent with the already existing photochemical models at all the heights we are able to sound in the UV wavelengths. This includes the survival time of the candidate, which should also be in agreement with the dynamical scales observed in the UV-markings \citep{titov2007}: for spatial scales of thousands of kilometers, we can identify clouds even after 4 days \citep{asl}. The planetary-scale dark markings created by the Kelvin wave responsible for the Y-feature (i.e. the dark phase of the wave) can be tracked as it distorts progressively up to 30 days \citep{rossow,peralta2015}.

In this work, we will only focus on the spectral characterization of the candidates. We will follow the short list provided by \cite{mills} of fewer than a dozen candidates, including the recent proposal by \cite{frandsen}. Table~2 shows some of these candidates together with the most early reference and the source of the spectral data to the absorption of each species. The last column ($\chi^2$) is the mean quadratic deviation from our results including their uncertainty to each candidate's absorption, computed below 0.6 $\mu$m

 \begin{table}
 \caption{Candidates for UV absorption}
 \label{table:candidates}
 \centering
\begin{tabular}{c c c c}
\hline 
{\bf Candidate}    & {\bf Reference}  & {\bf Spectral data} & {\bf $	\chi^2$}  \\ 
\hline 
S$_3$                        & \cite{toon}              & \cite{toon}              &  190 \\ 
S$_4$                        & \cite{toon}              & \cite{toon}              &  4$\times$10$^5$ \\ 
S$_8$ (25$^{\circ}$C)        & \cite{hapke}             & \cite{toon}              &  120 \\ 
S$_8$ (100$^{\circ}$C)       & \cite{hapke}             & \cite{toon}              &  76 \\ 
S$_2$O                       & \cite{hapkegraham}       & \cite{los2o}             &  14 \\ 
OSSO                         & \cite{frandsen}          & \cite{frandsen}          &  22 \\
SCl$_2$                      & \cite{krasnopolsky1986}  & \cite{colton}            &  57 \\  
(NH$_4$)$_2$S$_2$O$_5$       & \cite{titov}             & \cite{krasnopolsky1986}  &  164 \\  
NOHSO$_4$                    & \cite{watson}            & \cite{sill}              &  928 \\
Cl$_2$                       & \cite{pollack}           & \cite{pollack}           &  76 \\ 
FeCl$_3$                     & \cite{krasnopolsky1985}  & \cite{aoshima}           &  86 \\ 
C$_5$O$_5$H$_2$              & \cite{hartley}           & \cite{croconic}          &  1.5$\times$10$^4$ \\ 
\hline
\end{tabular}
\end{table}

Some of the candidates do not match at all the observed UV-absorption. Most notably, S$_4$ cannot account alone for the UV-marking, which is obvious as its absorption is centered at longer wavelengths. The croconic acid (C$_5$O$_5$H$_2$) proposed by \cite{hartley} and \cite{croconic} gives also a very bad spectral fit to our results, as happens with the nitrosulfuric acid proposed by \cite{watson}.

\begin{figure}[h]
\centering
\includegraphics[width=30pc]{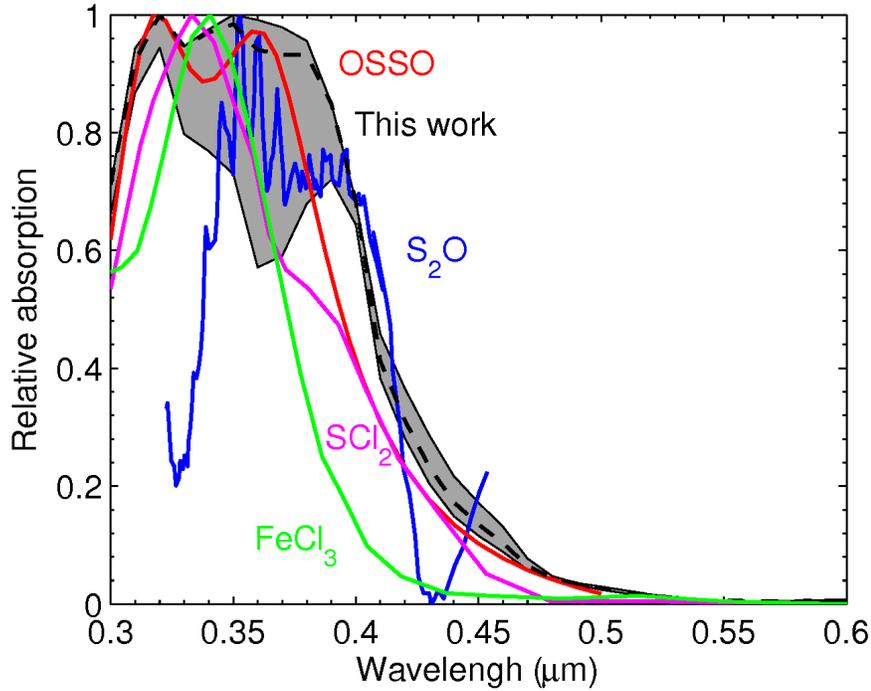}
\caption{A comparison of the relative absorption (in arbitrary units) of some of the candidates for the UV-absorber proposed so far with the model results obtained in this work. The grey area is used for our mode results, with the black dashed line being used for best-fitting values, and maximum and minimum absorption values being indicated with solid black lines. }
\label{candidates}
\end{figure}

It must be noted that most of the candidates, though not all, could account for the core absorption around 0.35 $\mu$m. The main problem is fitting the spectral slope between 0.4 $\mu$m and 0.5 $\mu$m with a single component. The best agreement is found for an irradiated version of S$_2$O from \cite{los2o} (whose application to the Venus problem might not be straightforward) and the recently proposed OSSO \citep{frandsen}. Other species have a too narrow absorption to be in agreement with our results. This happens for example with SCl$_2$, Cl$_2$ or FeCl$_3$.

Even though the average deviation is lower for S$_2$O, the spectral shape of our results resembles better that of OSSO and therefore they support this candidate if we are to attribute the absorption to just one absorber. As most of the candidates depart from our results around at 0.4 $\mu$m, it is tempting to argue in favor of a second absorber that could complement the absorber in this range. In such a case, S$_4$ would be an excellent candidate. Some arguments have been provided so far against sulfur compounds \citep{krasnopolsky2016} and in favour of FeCl$_3$ \citep{krasnopolsky2017} but, still, the spectral data are not close enough to support these species.

There are some uncertainties in this discussion. First, the assumed particle size has an effect on the width of the absorption band, with larger particles having a broader absorption. We have tested particle sizes from 0.1 $\mu$m to 0.5 $\mu$m and this would partially mitigate the deviation of OSSO or SCl$_2$ but it is not enough for FeCl$_3$. In any case, such model dependency must be highlighted. Second, the spectral data for all candidates are still very dispersed in the literature and often presented in a number of ways that prevent a straightforward comparison. Third, it would be desirable to have a better observational coverage of Venus spectrum from 0.35 $\mu$m to 0.5 $\mu$m. Fortunately, there is much information from VeX/VIRTIS that could be analyzed in the future and more recent observations from Akatsuki mission will help to correlate the variability of SO$_2$ and the UV-absorber \citep{lee2017}.

\subsection{The effect on energy budget}

The vertical distribution of the UV-absorber, together with its absorption spectrum, strongly influences the solar heating rates in Venus' mesosphere \citep{crisp}. This effect was measured with the Solar Flux Radiometer (LSFR) experiment on the large probe of the Pioneer Venus mission \citep{tomasko_heating}. Recent works have analyzed the radiative energy balance of Venus \citep{lee2015,haus2015, haus2016} and, in particular, the role of the UV-absorption in the solar heating.

While it is beyond the scope of this paper to provide a complete analysis of the solar heating rates, we can still discuss our results in terms of the parameterizations used so far to reproduce the shortwave range of the energy budget. There are essentially two aspects that must be addressed to this regard: vertical distribution of the absorber and its absorption spectrum. The rest of the atmospheric parameters also have an influence but can be taken as second order effects.

The vertical distribution of the UV-absorber has been taken in this work to be tied to the distribution of mode-1 particles, as done by \cite{crisp}. As shown in section~\ref{clouds}, our retrievals give an average value of 65$\pm$2 km for the base of mode-1 particles. This would be closer to or somewhat higher than the low UV-absorber model in \cite{haus2015,haus2016}. However, these works assume a scale height of 1 km, which is substantially lower than our $\sim$ 4 km scale height for mode-1. As this results in a cloud top, as previously defined, above 70 km, we argue that our results are closer to the nominal vertical distribution model by \cite{haus2015,haus2016}. It must be noted that our approach is unable to determine the vertical distribution of UV-absorber independently of mode-1. While other works have not been completely successful in confidently constraining the vertical distribution of the UV-absorber, they suggest that the unknown UV absorber may be more concentrated right below the cloud tops, and the bottom should be not deeper than 60 km \citep{lee2017}.

\begin{figure}[h]
\centering
\includegraphics[width=20pc]{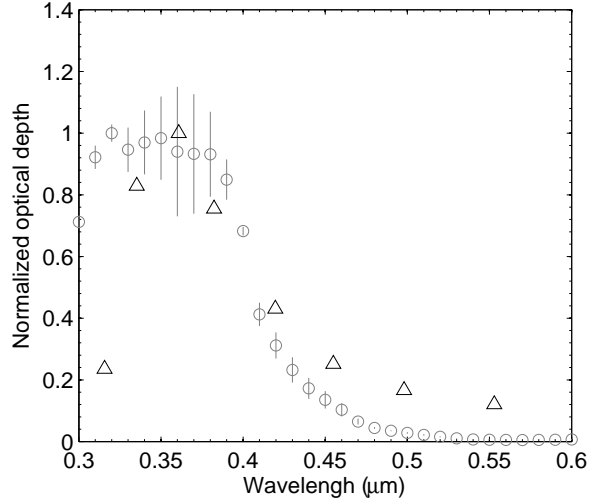}
\caption{Normalized optical thickness for the UV-absorption in this work (grey circles with error bars) compared with parameterization by \cite{haus2016} (black triangles).}
\label{haus}
\end{figure}

The second aspect of interest would be the spectral shape of the absorption itself. Here we will compare our results with those from \cite{haus2016}, shown in Fig.~\ref{haus}. The optical thickness is given normalized to the maximum value, to make it independent of the assumption of UV-absorber being attached to mode-1 particles. In this simple comparison we find that \cite{haus2016} absorption is in general agreement with our results for the core of the absorption bands but it is displaced towards red wavelengths. Absorption at 0.32 $\mu$m is significantly lower than ours but it is higher at 0.5 $\mu$m.

If we convolve each curve with the solar flux we find that the UV-absorber by \cite{haus2016} blocks a 40\% more flux than the one in our work, by direct absorption. This could result in an overestimation of the solar heating rates at these wavelengths.  However, the solar flux scattered by each model should also be taken into account in order to make an accurate determination of the impact of our results in the solar heating rates. In fact, \cite{lee2015} showed models should consider important factors such as the vertical variations of the cloud top altitude (raising the cloud top from 67 to 70 km can increase the heating rate at the cloud tops about a 50\%) or the horizontal distribution of the UV absorber as they are responsible for about a half of the total solar heating at the cloud tops.

\newpage
\section{Conclusions} \label{conclusions}

In this work, we have analyzed spectra taken during the MESSENGER spacecraft's second Venus flyby on June 5, 2007. In particular, the spectra were taken by the instrument MASCS that covers from the near ultraviolet (0.3 $\mu$m) to the near infrared (1.49 $\mu$m). Such spectra have been modeled with a radiative transfer model that is able to fit the observed radiance as a function of wavelength using as free parameters the vertical distribution of Venusian particles and the imaginary refractive index of the UV-absorber.

Our results show an Equatorial atmosphere with very homogeneous properties (particularly cloud tops and total particle density), whose cloud tops are located at 75$\pm$2 km, in good agreement with previous works. Our results are dominated by the particles located at 60 km or above, with little sensitivity below that.

The imaginary refractive index of the UV-absorber is found to be blue-shifted with respect to previous works, centered at 0.34$\pm$0.03 $\mu$m with a full width half maximum of 0.14$\pm$0.01 $\mu$m if we assume a gaussian shape for the absorption band. When comparing the spectral shape of the absorption with candidates proposed so far, we find that there is a better correlation with sulfur-bearing compounds such as S$_2$O or S$_2$O (in the form of OSSO $cis-$ and $trans-$ isomers), with SO$_2$ as a source, if we accept that the UV-absorption is produced by a single compound. The change in the UV-absorption may have an influence in the determination of the solar heating rates for Venus atmosphere that deserve future research.

The identification of the UV-absorber is a problem far from being solved. While this works supports the spectral similarity of the retrieved values with disulfur dioxide, it must be noted that the vertical distribution assumed here is not in complete agreement with the profiles computed by \cite{frandsen} . A more recent work \citep{krasnopolsky2018} also shows the weaknesses of this explanation in view of state of the art photochemical models of Venus' atmosphere.

As future work, there are essentially two aspects that should be investigated. Laboratory spectra of already proposed or new candidates for the UV absorption at the conditions of the upper Venus atmosphere (temperature, pressure and solar radiation) are required. Mid- to high-resolution Venus spectra would also be welcome, particularly if they provide spatial resolution to separate regions with higher and lower UV absorption. The wavelength range of 0.4--0.5 $\mu$m is of the highest interest, as it may provide very useful constraints on the nature and composition of the UV-absorption in Venus atmosphere. High-resolution spectra would also help to separate gaseous absorption from broader condensed matter absorption, thus constraining the physical state of the UV absorber.

\acknowledgments
This work was supported by the Spanish MICIIN projects AYA2015-65041-P (MINECO/FEDER, UE), Grupos Gobierno Vasco IT-765-13, and UFI11/55 from UPV/EHU. S.P.-H. acknowledges support from the Jose Castillejo Program funded by Ministerio de Educaci\'{o}n, Cultura y Deporte , Programa Nacional de Movilidad de Recursos Humanos del Plan Nacional de I-D+i 2008-2011.  J.P. acknowledges JAXA's International Top Young Fellowship (ITYF). MESSENGER MASCS VIRS calibrated data are publicly available through NASA Planetary Data System (http://pds-geosciences.wustl.edu/missions/-messenger/mascs.htm). The data supporting the
figures is available at http://www.ajax.ehu.eus/sph/papers/2017JE005406/. Access to code NEMESIS (http://users.ox.ac.uk-/~atmp0035/nemesis.html) is available upon request from Patrick Irwin (patrick.irwin@physics.ox.ac.uk).




\listofchanges


\begin{thebibliography}{}

\bibitem[{\textit{Aoshima et~al.}(2013)}]{aoshima}
Aoshima, H., K.~Satoh, T.~Umemura, and M.~Kamigaito (2013), A simple combination of higher-oxidation-state FeX$_3$ and phosphine or amine ligand for living radical polymerization of styrene, methacrylate, and acrylate, \textit{Polymer Chemistry}, \textit{4}, 3554--3562.

\bibitem[{\textit{Arney et~al.}(2014)}]{arney}
Arney, G., V.~Meadows, D.~Crisp, S.J.~Schmidt, J.~Bailey, and T.~Robinson (2014), Spatially resolved measurements of H$_2$O, HCl, CO, OCS, SO$_2$, cloud opacity, and acid concentration in the Venus near-infrared spectral windows, \textit{Journal of Geophys. Res.}, \textit{119}, 1860--1891.

\bibitem[{\textit{Barstow et~al.}(2012)}]{barstow}
Barstow, J.K., C.C.C.~Tsang, C.F.~Wilson, P.G.J.~Irwin, K.~McGouldrick, P.~Drossart, G.~Piccioni, and S.~Tellmann (2012), Models of the global cloud structure on Venus derived from Venus Express observations, \textit{Icarus}, \textit{217}, 542--560.
 it must be noted that 

\bibitem[{\textit{Belton et~al.}(1991)}]{belton}
Belton, M.J.S., P.J.~Gierasch, M.D.~Smith, P.~Helfenstein, P.J.~Schinder, J.B.~Pollack, K.A.~Rages, A.P.~Ingersoll, K.P.~Klaasen, J.~Veverka, C.D.~Anger, M.H.~Carr, C.R.~Chapman, M.E.~Davies, F.P.~Fanale, R.~Greeley, R.~Greenberg, J.W.~Head~III, D.~Morrison, G.~Neukum, and C.B.~Pilcher (1991), Images from Galileo of the Venus cloud deck, \textit{Science}, \textit{253}, 1531--1536.

\bibitem[{\textit{Belyaev et~al.}(2012)}]{belyaev2012}
Belyaev, D.~A., F.~Montmessin, J.-L.~Bertaux, A.~Mahieux, A.A.~Fedorova, O.I.~Korablev, E.~Marcq, Y.L.~Yung, and X.~Zhang (2012), Vertical profiling of SO$_2$ and SO above Venus' clouds by SPICAV/SOIR solar occultations, \textit{Icarus}, \textit{217}, 740--751.

\bibitem[{\textit{Belyaev et~al.}(2017)}]{belyaev2017}
Belyaev, D.~A., D.G.~Evdokimova, F.~Montmessin, J.-L.~Bertaux, O.I.~Korablev, A.A.~Fedorova, E.~Marcq, L.~Soret, and M.S.~Luginin (2017), Night side distribution of SO$_2$ content in Venus' upper mesosphere, \textit{Icarus}, \textit{294}, 58--71.

\bibitem[{\textit{Bertaux et~al.}(1996)}]{croconic}
Bertaux, J.-L., T.~Widemann, A.~Hauchecorne, V.I.~Moroz, A.P.~Ekonomov (1996), VEGA 1 and VEGA 2 entry probes: An investigation of local UV absorption (220--400 nm) in the atmosphere of Venus (SO$_2$ aerosols, cloud structure), \textit{Journal of Geophys. Res.}, \textit{101}, 12709--12745.

\bibitem[{\textit{Bertaux et~al.}(2016)}]{bertaux}
Bertaux, J.-L., I.V.~Khatuntsev, A.~Hauchecorne, W.J.~Markiewicz, E.~Marcq, S.~Lebonnois, M.~Patsaeva, A.~Turin, A.~Fedorova (2016), Influence of Venus topography on the zonal wind and UV albedo at cloud top level: The role of stationary gravity waves, \textit{Journal of Geophys. Res.}, \textit{121}, 1087--1101.

\bibitem[{\textit{Blackie et~al.}(2011)}]{so2}
Blackie, D., R.~Blackwell-Whitehead, G.~Stark, J.C.~Pickering, P.L.~Smith, J.~Rufus, and A.P.~Thorne (2011), High-resolution photoabsorption of SO$_2$ at 198K from 213 to 325 nm, \textit{Journal of Geophys. Res.}, \textit{116}, E03006.

\bibitem[{\textit{Boyer \& Camichel}(1961)}]{boyer}
Boyer, C., and H.~Camichel (1961), Observations photographiques de la plan{\`e}te Venus, \textit{Ann. Astrophys.}, \textit{24}, 531--535.

\bibitem[{\textit{de Bergh et~al.}(2006)}]{deberg} 
de Bergh, C., V.I.~Moroz, F.W.~Taylor, D.~Crisp, B.~B\'{e}zard, and L.V.~Zasova (2006), The composition of the atmosphere of Venus below 100 km altitude: An overview, \textit{Planet. and Space Sci.}, \textit{54}, 1389--1397.

\bibitem[{\textit{Clouthier}(1987)}]{s2o}
Clouthier, D.J. (1987), The Visible Absorption Spectrum of S$_2$O, \textit{Journal of Mol. Spec.}, \textit{124}, 179--184.

\bibitem[{\textit{Colina et~al.}(1996)}]{colina}http://www.ajax.ehu.eus/sph/papers/2017JE005406/
Colina, L., R.C.~Bohlin, and F.~Castelli (1996), The 0.12-2.5 $\mu$m absolute flux distribution of the Sun for comparison with solar analog stars, \textit{The Astronomical Journal}, \textit{112}, 307--315.

\bibitem[{\textit{Colton et~al.}(1974)}]{colton}
Colton, R.~J., and J.~W., Rayne (1974), Photoelectron and electronic absorption spectra of SCl$_2$, S$_2$Cl$_2$, S$_2$Br$_2$ and (CH$_3$)$_2$S$_2$, \textit{Journal of Electron Spectroscopy and Related Phenomena}, \textit{3}, 345--357.

\bibitem[{\textit{Crisp}(1986)}]{crisp}
Crisp, D. (1986), Radiative forcing of the Venus mesosphere: I. Solar fluxes and heating rates, \textit{Icarus}, \textit{67}, 484--514.

\bibitem[{\textit{Crisp et~al.}(1986)}]{crispnir}
Crisp, D., W.M.~Sinton, K.-W.~Hodapp, B.~Ragent, F.~Gerbault, J.H.~Goebel, R.G.~Probst, D.A.~Allen, K.~Pierce, and K.R.~Stapelfeldt (1986), The nature of the near-infrared features in the Venus night side, \textit{Science}, \textit{246}, 506--509.

\bibitem[{\textit{Ehrenheich et~al.}(2012)}]{venus_transiting}
Ehrenheich, D., A.~Vidal-Madjar, T.~Widemann, G.~Gronoff, P.~Tanga, M.~Barth{\'e}lemy, J.~Lilensten, A.~Lecavelier~des~Etangs, and L.~Arnold (2012), Transmission spectrum of Venus as a transiting exoplanet, \textit{Astronomy \& Astrophysics}, \textit{537}, L2.

\bibitem[{\textit{Esposito et~al.}(1983)}]{venusI}
Esposito, L.~W., R.~G.~Knollenberg, M.~I.~Marov, O.~B.~Toon, R.~P.~Turco (1983), The clouds are hazes of Venus, in \textit{Venus}, {Hunten}, D.~M. and {Colin}, L. and {Donahue}, T.~M. and {Moroz}, V.~I. (eds.), University of Arizona Press, Tucson, AZ, pp. 484--564

\bibitem[{\textit{Esposito et~al.}(1997)}]{venusII}
Esposito, L.~W., J.-L.~Bertaux,V.~Krasnopolsky, V.I.~Moroz, V. I., and L.V.~Zasova (1997), Chemistry of Lower Atmosphere and Clouds, in \textit{Venus II: Geology, Geophysics, Atmosphere, and Solar Wind Environment}, S.W. Bougher, D.M. Hunten, and R.J. Philips (eds.), University of Arizona Press, Tucson, AZ, pp. 415--458

\bibitem[{\textit{Fedorova et~al.}(2008)}]{fedorova}
Fedorova, A., O.~Korablev, A.-C.~Vandaele, J.-L.~Bertaux, D.~Belyaev, A.~Mahieux, E.~Neefs, W.V.~Wilquet, R.~Drummond, F.~Montmessin, and E.~Villard (2008), HDO and H$_2$O vertical distributions and isotopic ratio in the Venus mesosphere by Solar Occultation at Infrared spectrometer on board Venus Express, \textit{Journal of Geophys. Res.}, \textit{113}, E00B22.

\bibitem[{\textit{Frandsen et~al.}(2016)}]{frandsen}
Frandsen, B.N., P.O.~Wennberg, and H.G. Kjaergaard (2016), Identification of OSSO as a near-UV absorber in the Venusian atmosphere, \textit{Geophys. Res. Lett.}, \textit{43}, 11,146--11,155

\bibitem[{\textit{Gao et~al.}(2013)}]{gao}
Gao, P., X.~Zhang, D.~Crisp, C.G.~Bardeen, and Y.L.~Yung (2013), Bimodal Distribution of Sulfuric Acid Aerosols in the Upper Haze of Venus, \textit{Icarus}, \textit{231}, 83--98

\bibitem[{\textit{Garc\'{i}a-Mu\~{n}oz \& Mills}(2012)}]{agm2012}
Garc\'{i}a-Mu\~{n}oz, A., and F.P.~Mills (2012), The June 2012 transit of Venus. Framework for interpretation of observations, \textit{Astronomy \& Astrophys.}, \textit{547}, A22

\bibitem[{\textit{Garc\'{i}a-Mu\~{n}oz et~al.}(2013)}]{agm}
Garc\'{i}a-Mu\~{n}oz, A., P.~Wolkenberg, A.~S\'{a}nchez-Lavega, R.~Hueso, and I.~Garate-Lopez (2013), A model of scattered thermal radiation for Venus from 3 to 5 $\mu$m, \textit{Planet. and Space Sci.}, \textit{81}, 65--73

\bibitem[{\textit{Garc\'{i}a-Mu\~{n}oz et~al.}(2014)}]{agm_glory}
Garc\'{i}a-Mu\~{n}oz, A., S. P'{e}rez-Hoyos, and A.~S\'{a}nchez-Lavega (2014), Glory revealed in disk-integrated photometry of Venus, \textit{Astronomy \& Astrophys. Lett.}, \textit{566}, L1

\bibitem[{\textit{Grinspoon et~al.}(1993)}]{nims}
Grinspoon, D.H., J.B.~Pollack, B.R.Stamnes, K., Tsay, S.-C., Jayaweera, K., Wiscombe, W., R.W.~Carlson, L.W.~Kamp, K.H.~Baines, Th.~Encrenaz, and F.W.~Taylor (1993), Probing Venus cloud structure with Galileo NIMS, \textit{Planet. and Space Sci.}, \textit{41}, 515--542

\bibitem[{\textit{Hansen \& Hovenier}(1974)}]{venuspol}
Hansen, J.E., and J.W.~Hovenier (1974), Interpretation of the Polarization of Venus, \textit{Journal of the Atmos. Sci.}, \textit{31}, 1137--1160.		

\bibitem[{\textit{Hansen \& Travis}(1974)}]{hansentravis}
Hansen, J.E., and L.D. Travis (1974), \textit{Space Sci. Rev.}, \textit{16}, 527--610. 

\bibitem[{\textit{Hapke \& Nelson}(1975)}]{hapke}
Hapke, B., and R.~Nelson (1975), Evidence for an Elemental Sulfur Component of the Clouds from Venus Spectrophotometry , \textit{Journal of the Atmos. Sci.}, \textit{32}, 1212--1218.		

\bibitem[{\textit{Hapke \& Graham}(1989)}]{hapkegraham}
Hapke, B., and F.~Graham (1989), Spectral properties of condensed phases of disulfur monoxide, polysulfur oxide and irradiated sulfur, \textit{Icarus}, \textit{79}, 47--55.		

\bibitem[{\textit{Hartley et~al.}(1989)}]{hartley}
Hartley, K.~K., and A.~R. Wolff (1989), Croconic acid: and absorber in the Venus clouds?, \textit{Icarus}, \textit{77}, 382--390.		

\bibitem[{\textit{Haus et~al.}(2015)}]{haus2015}
Haus, R., D.~Kappel, and G.~Arnold (2015), Radiative heating and cooling in the middle and lower atmosphere of Venus and responses to atmospheric and spectroscopic parameter variations, \textit{Planet. and Space Sci.}, \textit{117}, 262--294.		

\bibitem[{\textit{Haus et~al.}(2016)}]{haus2016}
Haus, R., D.~Kappel, S.~Tellmann, G.~Arnold, G.~Piccioni, P.~Drossart, and B.~H\"{a}usler (2016), Radiative energy balance of Venus based on improved models of the middle and lower atmosphere, \textit{Icarus}, \textit{272}, 178--205.		

\bibitem[{\textit{Hawkins et~al.}(2007)}]{mdis}
Hawkins, S.E., III, J. Boldt, E.H. Darlington, R. Espiritu, R.E. Gold, B. Gotwols, M. Grey, C. Hash, J. Hayes, S. Jaskulek, C. Kardian, M. Keller, E. Malaret, S.L. Murchie, P. Murphy, K. Peacock, L. Prockter, A. Reiter, M.S. Robinson, E. Schaefer, R. Shelton, R. Sterner, H. Taylor, T. Watters, and B. Williams (2007), The Mercury Dual Imaging System on the MESSENGER spacecraft, \textit{Space Science Reviews}, \textit{131}, 247--338.

\bibitem[{\textit{Ignatiev et~al.}(2009)}]{ignatiev}
Ignatiev, N.I., D.V.~Titov, G.~Piccioni, P.~Drossart, W.J.~Markiewicz, V.~Cottini, Th.~Roatsch, M.~Almeida, and N.~Manoel (2009), Altimetry of the Venus cloud tops from the Venus Express observations, \textit{Journal of Geophys. Res.}, \textit{114}, E00B43.

\bibitem[{\textit{Imai et~al.}(2016)}]{imai}
Imai, M., Y.~Takahashi, M.~Watanabe, T.~Kouyama, S.~Watanabe, S.~Gouda, and Y.~Gouda (2016), Ground-based observations of the cyclic nature and temporal variability of planetary-scale UV features at the Venus cloud top level, \textit{Icarus}, \textit{278}, 204--214.

\bibitem[{\textit{Irwin et~al.}(2008)}]{nemesis}
Irwin, P.G.J., N.A.~Teanby, R.~de Kok, L.N.~Fletcher, C.~Howett, C.C.C.~Tsang, C.F.~Wilson, S.B.~Calcutt, C.A.~Nixon, P.D.~Parrish (2008), The NEMESIS planetary atmosphere radiative transfer and retrieval tool, \textit{Journal of Quant. Spec. and Rad. Trans.}, \textit{109}, 1136--1150.

\bibitem[{\textit{Irwin et~al.}(2015)}]{improvement}
Irwin, P.G.J., D.S.~Tice, L.N.~Fletcher, J.K.~Barstow, N.A.~Teanby, G.S.~Orton, and G.R.~Davis (2015), Reanalysis of Uranus' cloud scattering properties from IRTF/SpeX observations using a self-consistent scattering cloud retrieval scheme,\textit{Journal of Quant. Spec. and Rad. Trans.}, \textit{109}, 1136--1150.

\bibitem[{\textit{Khatuntsev et~al.}(2013)}]{khatuntsev}
Khatuntsev, I.~V., M.V.~Patsaeva, D.V.~Titov, N.I.~Ignatiev, A.V.~Turin, S.S.~Limaye, W.J.~Markiewicz, M.~Almeida, Th.~Roatsch, and R.~Moissl (2013), Cloud level winds from the Venus Express Monitoring Camera imaging, \textit{Icarus}, \textit{226}, 140--158.

\bibitem[{\textit{Knollenberg \& Hunten}(1980)}]{knollenberg}
Knollenberg, R.G., and D.M.~Hunten (1980), The microphysics of the clouds of Venus: results from the Pioneer Venus particle size spectrometer experiment, \textit{Journal of Geophys. Res.}, \textit{85}, 8039--8058.

\bibitem[{\textit{Krasnopolsky}(1985)}]{krasnopolsky1985}
Krasnopolsky, V.A. (1985), Chemical composition of Venus clouds, \textit{Planet. Space Sci.}, \textit{33}, 109--117.

\bibitem[{\textit{Krasnopolsky}(1986)}]{krasnopolsky1986}
Krasnopolsky, V.A. (1986), \textit{Photochemistry of the atmosphere of Mars and Venus}, Springer-Verlag, Berlin.

\bibitem[{\textit{Krasnopolsky}(2006)}]{krasnopolsky2006}
Krasnopolsky, V.A. (2006), Chemical composition of Venus atmosphere and clouds: Some unsolved problems, \textit{Planet. and Space Sci.}, \textit{54}, 1352--1359.

\bibitem[{\textit{Krasnopolsky}(2016)}]{krasnopolsky2016}
Krasnopolsky, V.A. (2016), Sulfur aerosol in the clouds of Venus, \textit{Icarus}, \textit{274}, 33--36.

\bibitem[{\textit{Krasnopolsky}(2017)}]{krasnopolsky2017}
Krasnopolsky, V.A. (2017), On the iron chloride aerosol in the clouds of Venus, \textit{Icarus}, \textit{286}, 134--137.

\bibitem[{\textit{Krasnopolsky}(2018)}]{krasnopolsky2018}
Krasnopolsky, V.A. (2018), Disulfur dioxide and its near-UV absorption in the photochemical model of Venus atmosphere, \textit{Icarus}, \textit{299}, 294--299.

\bibitem[{\textit{Krasnopolsky et~al.}(2013)}]{krasnopolsky2013}
Krasnopolsky, V.A., D.A.~Belyaev, I.E.~Gordon, G.~Li, L.S.~Rothman (2013), Observations of D/H ratios in H2O, HCl, and HF on Venus and new DCl and DF line strengths, \textit{Icarus}, \textit{224}, 57--65.

\bibitem[{\textit{Lee et~al.}(2012)}]{lee2012}
Lee, Y.J., D.V.~Titov, S.~Tellmann, A.~Piccialli, N.~Ignatiev, M.~Pl\"{a}tzold, B.~H\"{a}usler, G.~Piccioni, and P.~Drossart (2012), Vertical structure of the Venus cloud top from the VeRa and VIRTIS observations onboard Venus Express, \textit{Icarus}, \textit{217}, 599--609.

\bibitem[{\textit{Lee et~al.}(2015a)}]{lee2015}
Lee, Y.J., T.~Imamura, S.E.~Schr\"{o}eder, and E.~Marcq (2015), Long-term variations of the UV contrast on Venus observed by the Venus Monitoring Camera on board Venus Express, \textit{Icarus}, \textit{253}, 1--15.

\bibitem[{\textit{Lee et~al.}(2015b)}]{lee2015b}
Lee, Y.J., D.V.~Titov, N.I.~Ignatiev, S.~Tellmann, M.~P\"atzold, and G. Piccioni (2015), The radiative forcing variability caused by the changes of the upper cloud vertical structure in the Venus mesosphere, \textit{Planet. and Space Sci.}, \textit{113--114}, 298--308.

\bibitem[{\textit{Lee et~al.}(2017)}]{lee2017}
Lee, Y.J., A.~Yamazaki, T.~Imamura, M.~Yamada, S.~Watanabe, T.M.~Sato, K.~Ogohara, G.L.~Hashimoto, and S.~Murakami (2017), Scattering properties of the Venusian clouds observed by the UV imager on board Akatsuki, \textit{The Astronomical Journal}, \textit{154}, 44--60.

\bibitem[{\textit{Lo et~al.}(2003)}]{los2o}
Lo, W.~J., Y.-J. Wu, and Y.-P. Lee (2003), Ultraviolet absorption spectrum of cyclic S$_2$O in solid Ar, \textit{J. Phys. Chem. A}, \textit{107}, 6944--6947.

\bibitem[{\textit{Luginin et~al.}(2016)}]{luginin}
Luginin, M., A.~Fedorova, D.~Belyaev, F.~Montmessin, V.~Wilquet, O.~Korablev, J.-L.~Bertaux, and A.C.~Vandaele (2016), Aerosol properties in the upper haze of Venus from SPICAV IR data, \textit{Icarus}, \textit{277}, 154--170.

\bibitem[{\textit{Marcq et~al.}(2011)}]{marcq2011}
Marcq, E., D.~Belyaev, F.~Montmessin, A.~Fedorova, J.-L.~Bertaux, A.C.~Vandaele, and E.~Neefs (2011), An investigation of the SO$_2$ content of the venusian mesosphere using SPICAV-UV in nadir mode, \textit{Icarus}, \textit{211}, 58--69.

\bibitem[{\textit{Marcq et~al.}(2013)}]{marcq2013}
Marcq, E., J.-L.~Bertaux, F.~Montmessin, and  D.~Belyaev (2013), Variations of sulphur dioxide at the cloud top of Venus's dynamic atmosphere, \textit{Nat. Geosci.}, \textit{6}, 25--28.

\bibitem[{\textit{Markiewicz et~al.}(2014)}]{glory}
Markiewicz, W.J., E.~Petrova, O.~Shalygina, M.~Almeida, D.V.~Titov, S.S.~Limaye, N.~Ignatiev, T.~Roatsch, and K.D.~Matz (2014), Glory on Venus cloud tops and the unknown UV absorber, \textit{Icarus}, \textit{234}, 200--203.

\bibitem[{\textit{McClintock \& Lankton}(2007)}]{mascs}
McClintock, W.E., and M.R.~Lankton (2007), The Mercury Atmospheric and Surface Composition Spectrometer for the MESSENGER Mission, \textit{Space Sci. Rev.}, \textit{131}, 481--521.

\bibitem[{\textit{McGouldrick \& Toon}(2007)}]{holes}
McGouldrick, K., and O.B.~Toon (2007), An investigation of possible causes of the holes in the condensational Venus cloud using a microphysical cloud model with a radiative-dynamical feedback, \textit{Icarus}, \textit{191}, 1--24.

\bibitem[{\textit{McNutt et~al.}(2008)}]{messenger}
McNutt, R.L., S.C.~Solomon, D.G.~Grant, E.J.~Finnegan, P.D.~Bedini, and the MESSENGER Team (2008), The MESSENGER mission to Mercury: Status after the Venus flybys, \textit{Acta Astronautica}, \textit{63}, 68--73.

\bibitem[{\textit{Mills et~al.}(2007)}]{mills}
Mills, F.~P., L.~W. Esposito, and Y.~L. Yung (2007), Atmospheric composition, chemistry and clouds, in \textit{Exploring Venus as a terrestrial planet}, American Geophysical Union, Washington, DC.

\bibitem[{\textit{Minnaert}(1941)}]{minnaert}
Minnaert, M. (1941), The reciprocity principle in lunar photometry, \textit{Astrophys. J.}, \textit{93}, 403--410.
A. García Muñoz, P. Wolkenberg, A. Sánchez-Lavega, R. Hueso, I. Garate-Lopez

\bibitem[{\textit{Molaverdikhani et~al.}(2012)}]{molaverdikhani}
Molaverdikhani, K., K.~McGouldrick, and L.W.~Esposito (2012), The abundance and vertical distribution of the unknown ultraviolet absorber in the venusian atmosphere from analysis of Venus Monitoring Camera images, \textit{Icarus}, \textit{217}, 648--660.

\bibitem[{\textit{Murray et~al.}(1974)}]{murray}
Murray, B.C., M.J.S.~Belton, G.E.~Danielson, M.E.~Davies, D.~Gault, B.~Hapke, B.~O'Leary, R.G.~Strom, V.~Suomi, and N.~Trask (1974), Atmospheric motion and structure from Mariner 10 pictures, \textit{Science}, \textit{183}, 1307--1315.

\bibitem[{\textit{Murtagh \& Heck}(1987)}]{pca}
Murtagh, F. and Heck, A. (1987), Multivariate Data Analysis, Reidel, Dordrecht, Netherlands.

\bibitem[{\textit{Na \& Esposito}(1997)}]{na}
Na, C.Y., and L.W.~Esposito (1997), Is Disulfur Monoxide a Second Absorber on Venus?, \textit{Icarus}, \textit{125}, 364--368.

\bibitem[{\textit{Palmer \& Williams}(1975)}]{palmer}
Palmer, K.F., and D.~Williams (1975), Optical constants of sulfuric Acid; application to the clouds of venus?, \textit{Appl. Opt.}, \textit{14}, 208--219.

\bibitem[{\textit{Peralta et~al.}(2015)}]{peralta2015}
Peralta, J., A.~S\'{a}nchez-Lavega, M.A.~L\'{o}pez-Valverde, D.~Luz, and P.~Machado (2015). Venus' major cloud feature as an equatorially trapped wave distorted by the wind, \textit{Geophys. Res. Lett.}, \textit{42}, 705--711.

\bibitem[{\textit{Peralta et~al.}(2017a)}]{peralta}
Peralta, J., Y.J.~Lee, K.~McGouldrick, H.~Sagawa, A.~S\'{a}nchez-Lavega, T.~Imamura, T.~Widemann, and M.~Nakamura (2017). Overview of useful spectral regions for Venus: and update to encourage observations complementary to the Akatsuki mission, \textit{Icarus}, \textit{288}, 235--239.

\bibitem[{\textit{Peralta et~al.}(2017b)}]{peraltaMDIS}
Peralta, J., Y.J.~Lee, R. Hueso, R.T.~Clancy, B.J.~Sandor, A.~S\'{a}nchez-Lavega, E.~Lellouch, M.~Rengel, P.~Machado, M.~Omino, A.~Piccialli, T.~Imamura, T.~Horinouchi, S.~Murakami, K.~Ogohara, D.~Luz and D.~Peach (2017). Venus's Winds and Temperatures during the Messenger's flyby: an approximation to a three-dimensional instantaneous state of the atmosphere, \textit{Geophys. Res. Lett.}, \textit{44}, 3907--3915.

\bibitem[{\textit{Petrova et ~al.}(2015)}]{petrova}
Petrova, E.V., O.S.~Shalygina, and W.J.~Markiewicz (2015), UV contrasts and microphysical properties of the upper clouds of Venus from the UV and NIR VMC/VEx images, \textit{Icarus}, \textit{260}, 190--204.

\bibitem[{\textit{Pollack et~al.}(1980)}]{pollack}
Pollack, J.B., O.B.~Toon, R.C.~Whitten, R.~Boese, B.~Ragent, M.~Tomasko, L.~Esposito, L.~Travis, and D.~Wiedman (1980), Distribution and source of the UV absorption in Venus' atmosphere, \textit{Journal of Geophys. Res.}, \textit{85}, 8141--8150.

\bibitem[\textit{Pollack et~al.}(1993)]{pollack1993}
Pollack, J., Dalton, J., Grinspoon, D., Wattson, R., Freedman, R., Crisp, D., Allen, D., B\'{e}zard, B., de Bergh, C., Giver, L., (1993), Near-infrared light from Venus' nightside: A spectroscopic analysis. \textit{Icarus}, \textit{103}, 1--42.

\bibitem[{\textit{Rodgers}(2000)}]{rodgers}
Rodgers, C.D. (2000), Inverse Methods for Atmospheric Sounding, Theory and Practice, World Scientific, London.

\bibitem[{\textit{Ross}(1928)}]{ross}
Ross, F.E. (1928), Photographs of Venus, \textit{Astrophys. Journal}, \textit{68}, 57--92.

\bibitem[{\textit{Rossi et~al.}(2015)}]{rossi}
Rossi, L., E.~Marcq, F.~Montmessin, A.~Fedorova, D.~Stam, J.-L.~Bertaux, and O.~Korablev (2015), Preliminary study of Venus cloud layers with polarimetric data from SPICAV/VEx, \textit{Planet. Space Sci.}, \textit{113-114}, 159--168.

\bibitem[{\textit{Rossow et~al.}(1980)}]{rossow}
Rossow, W.B., A.D.~Del~Genio, S.S.~Limaye, and L.D.~Travis (1980), Cloud morphology and motions from Pioneer Venus images, \textit{Journal of Geophys. Res.}, \textit{85}, 8107--8128.

\bibitem[{\textit{Rothman et~al.}(2013)}]{hitran}
Rothman, L.S.,  I.E.  Gordon,  Y.  Babikov,  A.  Barbe,  D.Chris  Benner,  P.F.  Bernath,  et  al. (2013), The HITRAN 2012 molecular spectroscopic database, \textit{Journal of Quant. Spec. and Rad. Trans.}, \textit{130}, 4--50. 

\bibitem[{\textit{S\'{a}nchez-Lavega}(2011)}]{aslbook}
S\'{a}nchez-Lavega, A. (2011), An introduction to planetary atmospheres, \textit{CRC Press}, Boca Raton, FL, USA.

\bibitem[{\textit{S\'{a}nchez-Lavega et~al.}(2016)}]{asl}
S\'{a}nchez-Lavega, A., J.~Peralta, J.M.~G\'omez-Forellad, R.~Hueso, S.~P\'erez-Hoyos, I.~Mendikoa, J.F.~Rojas, T.~Horinouchi, Y.J.~Lee, S.~Watanabe (2016), Venus cloud morphology and motions from ground-based images at the time of the Akatsuki orbit insertion, \textit{Astrophys. Journal Lett.}, \textit{833}, L7
 
\bibitem[{\textit{Sato et~al.}(2014)}]{sato}
Sato, T.M., H.~Sagawa, T.~Kouyama, K.~Mitsuyama, T.~Satoh, S.~Ohtsuki, M.~Ueno, Y.~Kasaba, M.~Nakamura, and T.~Imamura (2014), Cloud top structure of Venus revealed by Subaru/COMICS mid-infrared images, \textit{Icarus}, \textit{243}, 386--399.

\bibitem[{\textit{Shalygina et~al.}(2015)}]{shalygina}
Shalygina, O., E.V.~Petrova, W.J.~Markiewicz, N.I.~Ignatiev, and E.V. Shalygin (2015), Optical properties of the Venus upper clouds from the data obtained by Venus Monitoring Camera on-board the Venus Express, \textit{Planet. and Space Sci.}, \textit{113--114}, 135--158.

\bibitem[{\textit{Sill et~al.}(1983)}]{sill}
Sill, G.~T. (1983), The clouds of Venus: Sulfuric Acid by the lead chamber process, \textit{Icarus}, \textit{53}, 10--17.

\bibitem[{\textit{Stamnes et~al.}(1988)}]{disort}
Stamnes, K., Tsay, S.-C., Jayaweera, K., Wiscombe, W. (1988), Numerically stable algorithm for discrete-ordinate-method radiative transfer in multiple scattering and emitting layered media, \textit{Applied Optics}, \textit{27}, 2502--2509.

\bibitem[{\textit{Takagi \& Iwagami}(2011)}]{takagi}
Takagi, S., and N.~Iwagami (2011), Contrast sources for the infrared images taken by the Venus mission AKATSUKI, \textit{Earth, Planets and Space}, \textit{63}, 435--442.

\bibitem[{\textit{Titov}(1983)}]{titov}
Titov,~D.~V. (1983), On the possibility of aerosol formation by the reaction between SO$_2$ and NH$_3$ in Venus, \textit{Cosmic Res.}, \textit{21}, 401.

\bibitem[{\textit{Titov et~al.}(2007)}]{titov2007}
Titov,~D.~V., F.W.~Taylor, H.~Svedhem, N.I.~Ignatiev, W.J.~Markiewicz, G.~Piccioni, and P.~Drossart (2007), Atmospheric structure and dynamics as the cause of ultraviolet markings in the clouds of Venus, \textit{Nature}, \textit{456}, 620--623.

\bibitem[{\textit{Titov et~al.}(2012)}]{titov2012}
Titov,~D.~V., W.J.~Markiewicz, N.I.~Ignatiev, L.~Song, S.S.~Limaye, A.~S\'anchez-Lavega, J.~Hesemann, M.~Almeida, T.~Roatsch, K.-D.~Matz, F.~Scholten, D.~Crisp, L.W.~Esposito, S.F.~Hviid, R.~Jaumann, H.U.~Keller, and R.~Moissl (2012), Morphology of the cloud tops as observed by the Venus Express Monitoring Camera, \textit{Icarus}, \textit{217}, 682--701.

\bibitem[{\textit{Tomasko et~al.}(1980)}]{tomasko}
Tomasko,~M.G., L.R.~Doose, P.H.~Smith, and A.P.~Odell  (1980), Measurements of the flux of sunlight in the atmosphere of Venus, \textit{Journal of Geophys. Res.}, \textit{85}, 8167--8186.

\bibitem[{\textit{Tomasko et~al.}(1985)}]{tomasko_heating}
Tomasko,~M.G., L.R.~Doose, and P.H.~Smith (1985), The absorption of solar energy and the heating rate in the atmosphere of Venus, \textit{Adv. Space Res.}, \textit{5}, 71--79.

\bibitem[{\textit{Toon et~al.}(1982)}]{toon}
Toon, O.B., R.P.~Turco, and J.B.~Pollack (1982), The ultraviolet absorber on Venus: Amorphous sulfur, \textit{Icarus}, \textit{51}, 358--373.

\bibitem[{\textit{Tsang et~al.}(2008)}]{tsang2008}
Tsang, C.C.C., P.G.J.~Irwin, F.W.~Taylor, and C.F.~Wilson (2008), A correlated-k model of radiative transfer in the near-infrared windows of Venus, \textit{Journal of Quant. Spec. and Rad. Trans.}, \textit{109}, 1118--1135.

\bibitem[{\textit{Tsang et~al.}(2010)}]{tsang2010}
Tsang, C.C.C., C.F.~Wilson. J.K.~Barstow, P.G.J.~Irwin, F.W.~Taylor, K. McGouldrick, G.~Piccioni, P.~Drossart, and H. Svedhem (2010), Correlations between cloud thickness and sub-cloud water abundance, \textit{Geophys. Res. Lett.}, \textit{37}, L02202.

\bibitem[{\textit{Vandaele et~al.}(2017)}]{vandaele}
Vandaele, A.~C., O.~Korablev, D.~Belyaev, S.~Chamberlain, D.~Evdokimova, T.~Encrenaz, L.~Esposito, K.~L.~Jessup, F.~Lef{\`e}vre, S.~Limaye, A.~Mahieux, E.~Marcq, F.P.~Mills, F.~Montmessin, C.D.~Parkinson, S.~Robert, T.~Roman, B.~Sandor, A.~Stolzenbach, C.~Wilson,  and V.~Wilquet (2017), CSulfur dioxide in the Venus atmosphere: I. Vertical distribution and variability, \textit{Icarus}, \textit{295}, 16--33.

\bibitem[{\textit{Watson et~al.}(1979)}]{watson}
Watson, A.~J., T.~M. Donahue, D.~H. Stedman, R.~G. Knollenberg, B.~Ragent, and J.~Blamont (1979), Oxides of Nitrogen and the clouds of Venus, \textit{Geophys. Res. Lett.}, \textit{6}, 743--746.

\bibitem[{\textit{Winick \& Stewart}(1980)}]{cloro}
Winick, J.R., and A.I.~Stewart (1980), Photochemistry of SO$_2$ in Venus' upper clouds, \textit{Journal of Geophys. Res.}, \textit{85}, 7849--7860.

\bibitem[{\textit{Zasova et~al.}(1981)}]{zasova1981}
Zasova, L.V., V.A.~Krasnopolsky, and V.I.~Moroz (1981), Vertical distribution of SO$_2$ in upper cloud layer of Venus and origin of the U.V.-absorption, \textit{Adv. Space Res.}, \textit{1}, 13--16.

\bibitem[{\textit{Zasova et~al.}(2007)}]{zasova2007}
Zasova, L.V., N.~Ignatiev, I.~Khatuntsev, and V.~Linkin (2007), Structure of the Venus atmosphere, \textit{Planet. and Space Sci.}, \textit{55}, 1712--1728.

\bibitem[{\textit{von Zahn \& Moroz}(1983)}]{vira}
von Zhan, U., and V.I.~Moroz (1983), Composition of the Venus atmosphere below 100 km, \textit{Adv. Space Res.}, \textit{5}, 173--195.

\end{thebibliography}
\end{document}